\documentclass[aps,longbibliography,pre,reprint,groupaddress,bibnotes]{revtex4-1}

\usepackage{dcolumn}% Align table columns on decimal point
\usepackage{bm, amssymb, color}% bold math
\usepackage{hyperref}% add hypertext capabilities
\usepackage{natbib}
\usepackage{tabulary}
\usepackage{multirow}
\usepackage{amsmath,soul}
\usepackage[export]{adjustbox}
\usepackage{float}
\usepackage[caption=false]{subfig}
\usepackage{array}
\usepackage{cases}
\usepackage{makecell}
\usepackage{physics}
\usepackage{yhmath}
\usepackage{enumitem}

\usepackage[colorinlistoftodos,shadow,textwidth=18mm]{todonotes}

\usepackage{soul,xcolor}
\setstcolor{red}

\usepackage[colorinlistoftodos,shadow,textwidth=18mm]{todonotes}

\begin{document}

\title{Diffusion Spreadability as a Probe of the Microstructure
of Complex Media Across Length Scales}

\author{Salvatore Torquato}
\email[]{Email: torquato@princeton.edu}
\affiliation{Department of Chemistry, Department of Physics,
Princeton Institute for the Science and Technology of Materials, and Program in Applied and Computational Mathematics, Princeton University, Princeton, New Jersey 08544, USA}

\begin{abstract}
Understanding time-dependent diffusion processes in multiphase media is of great importance in physics,
chemistry, materials science,  petroleum engineering and biology. Consider the time-dependent problem of mass transfer of a solute between two phases
and assume that the solute is initially distributed in one phase (phase 2) and absent from the other (phase 1). We desire the fraction of total solute present in phase 1 as a function of time, ${\cal S}(t)$, which we call the  {\it spreadability}, since it is a measure of the spreadability of diffusion information
as a function of time.
%Surprisingly, the consequences of this exact result, which dates back
%to early work by Prager, are unknown because it has yet to be understood fundamentally or applied in any meaningful way.
We derive exact direct-space formulas for ${\cal S}(t)$ in any Euclidean space dimension $d$ in terms
of the autocovariance function as well as  corresponding Fourier representations of ${\cal S}(t)$ in terms of the spectral density,  which are especially useful when scattering information is available experimentally or theoretically.
These are singular results because they are rare examples of
mass transport problems where exact solutions are possible. We  derive closed-form general formulas
for the short- and long-time behaviors of the spreadability  in terms of crucial small- and large-scale microstructural information, respectively. The long-time behavior of ${\cal S}(t)$ enables one
to distinguish the entire spectrum of microstructures that span from hyperuniform to nonhyperuniform media.  For hyperuniform media, disordered or not, we show that the ``excess" spreadability, ${\cal S}(\infty)-{\cal S}(t)$,
decays to its long-time behavior exponentially faster than that of any nonhyperuniform two-phase medium, the ``slowest" being
antihyperuniform media. The stealthy hyperuniform class is characterized by an excess spreadability  with the fastest decay rate among all translationally invariant microstructures. 
We obtain exact results for ${\cal S}(t)$ for a variety of specific
ordered and disordered model microstructures across  dimensions that span from hyperuniform to antihyperuniform
media. Moreover, we establish a remarkable connection between the spreadability and an outstanding problem in discrete geometry, namely, 
microstructures with ``fast" spreadabilities are also those that can be derived from efficient ``coverings" of space. 
We also identify heretofore unnoticed  remarkable links between the spreadability ${\cal S}(t)$ and
  NMR pulsed field gradient spin-echo amplitude as well as diffusion MRI measurements.
This investigation reveals that the time-dependent spreadability is a powerful, new dynamic-based figure of merit to probe and classify
the spectrum of possible microstructures of two-phase media across length scales.

\end{abstract}

\maketitle

\section{Introduction}

Interphase diffusion processes in heterogeneous media
are ubiquitous in a variety of contexts and applications, including  magnetic resonance imaging (MRI) \cite{We05},  geological media \cite{To02a,Sa03,Da13,Ta18}, 
biological cells \cite{Br79,Hof13},  and controlled drug delivery \cite{La81}.
%The physical properties of petroleum reservoir rock  formations are fundamentally controlled by the underlying complex microstructures
%of these heterogeneous media across length scales. Knowledge of their diffusion, flow and mechanical properties
%is of central importance in virtually all petroleum engineering applications and particularly significant in reservoir
%engineering \cite{Or02,Da13}. While these physical properties can be ascertained experimentally
%by retrieving physical samples and testing them in the laboratory,
%such methods can be tedious and often do not yield insights
%about the microstructures. The emergence of  enhanced recovery techniques in the field of oil and gas production 
%emphasizes the need for sophisticated mathematical and computational tools that are
%capable of modeling the complex microstructures and their physical properties.
%For example, X-ray microtomographic techniques, which provide 3D digitized images of rocks 
%at the pore scale in a nondestructive manner, can now be combined with computational 
%reconstruction techniques \cite{Ta13,Ta18} to produce a wide class of realistic 
%microstructures at will. Subsequently, such computer-generated
%microstructures can be structurally characterized and their physical properties 
%can be estimated via computer simulations \cite{Sa03,Wo03,Bi06,Van17,Ar18}.
In an unheralded paper published in 1963, Prager considered 
the time-dependent problem of mass transfer of solute between two phases
of a heterogeneous medium in three-dimensional Euclidean space $\mathbb{R}^3$ \cite{Pr63b}.
Phases 1 and 2 occupy volume fractions $\phi_1$ and $\phi_2$, respectively.
He assumed that a solute that is being transferred
from one phase to the other has the same diffusion
coefficient $D$ in each phase.  At $t = 0$, the solute is uniformly distributed
throughout phase 2, and completely absent from
phase 1. Prager desired  to calculate the fraction of the
total amount of solute present that has diffused into
phase 1 at time $t$, which we denote by ${\cal S}(t)$; see Fig. 1 for a schematic illustrating the spreadability phenomena for a special microstructure.
For two different microstructures at a given time $t$, the one with the larger value of ${\cal S}(t)$
spreads diffusion information more rapidly. For this reason, we henceforth call the time-dependent function ${\cal S}(t)$ the {\it spreadability}. Prager recognized that this problem can be solved exactly and found the following exact
direct-space solution in three dimensions:
\begin{equation}
{\cal S}(t)= \frac{1}{(4\pi D t)^{3/2}\, \phi_2} \int_{\mathbb{R}^3} [\phi_2-S_2({\bf r})] \exp[-r^2/(4Dt)] d{\bf r},
\label{0}
\end{equation}
where $S_2({\bf r})$ is the two-point probability function of phase 2 (defined in Sec. \ref{back}).
This is a singular result  because it represents one of the rare examples of
interphase mass transfer in two-phase random media where an
exact solution is possible only in terms of $\phi_2$ and $S_2$. Generally,
the effective properties of heterogeneous media are determined
not only by $\phi_2$ and $S_2$ but  all of the corresponding high-order correlation functions, which constitutes
a countably infinite set \cite{To02a}.

\begin{figure*}[bt]
\includegraphics[width=2.1in,keepaspectratio,clip=]{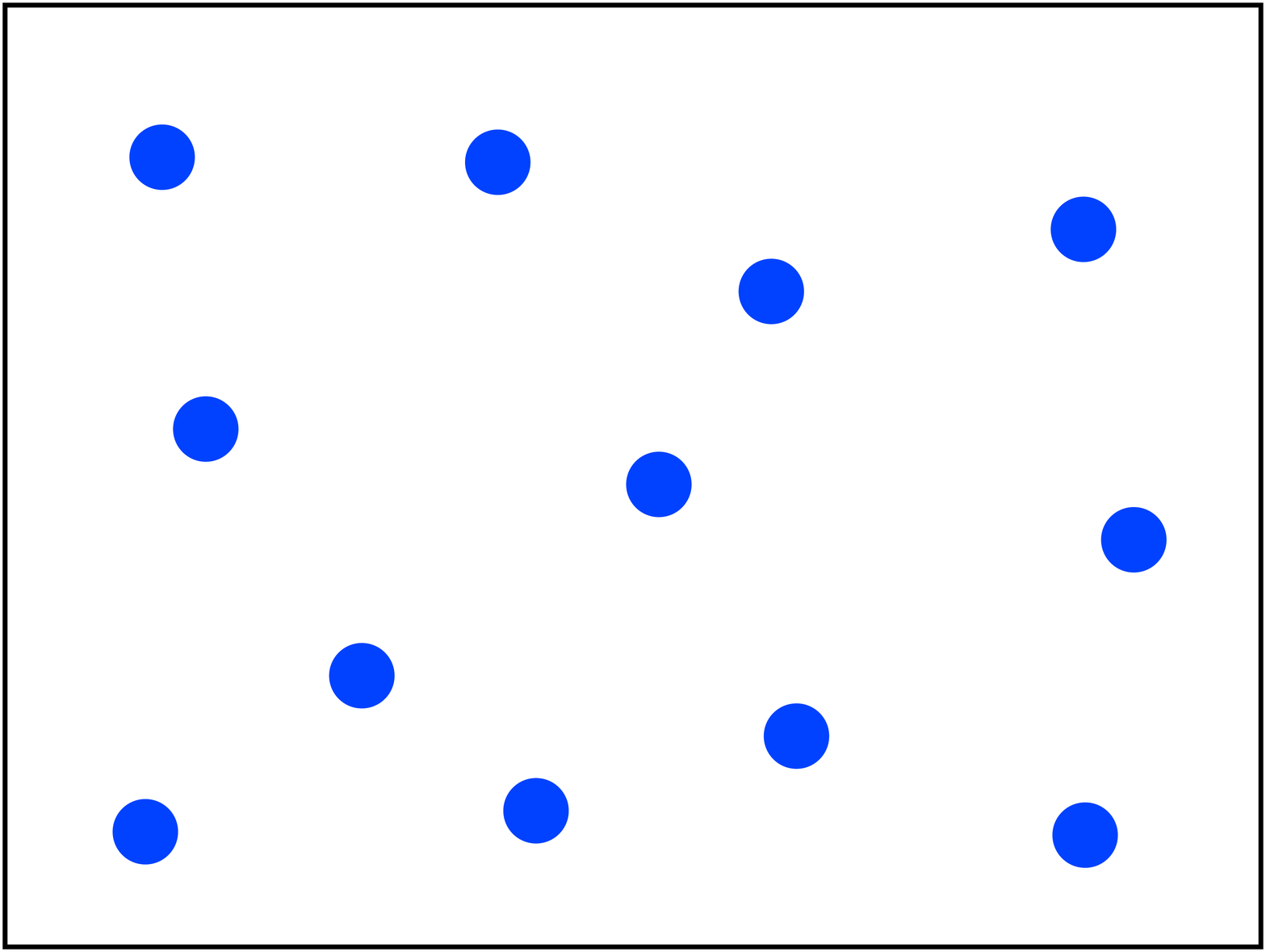}
\hspace{0.2in}\includegraphics[width=2.1in,keepaspectratio,clip=]{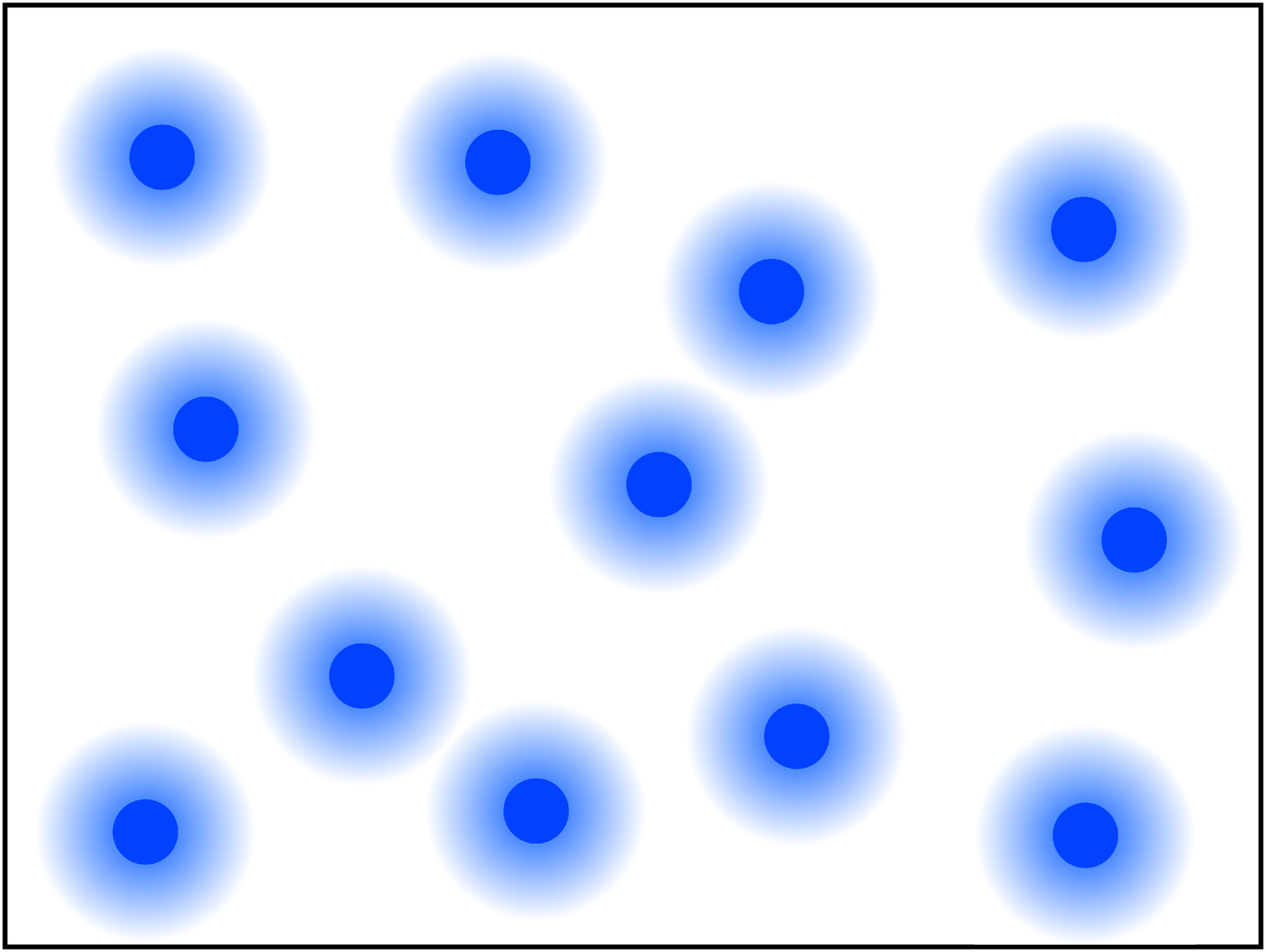}
\hspace{0.2in}\includegraphics[width=2.1in,keepaspectratio,clip=]{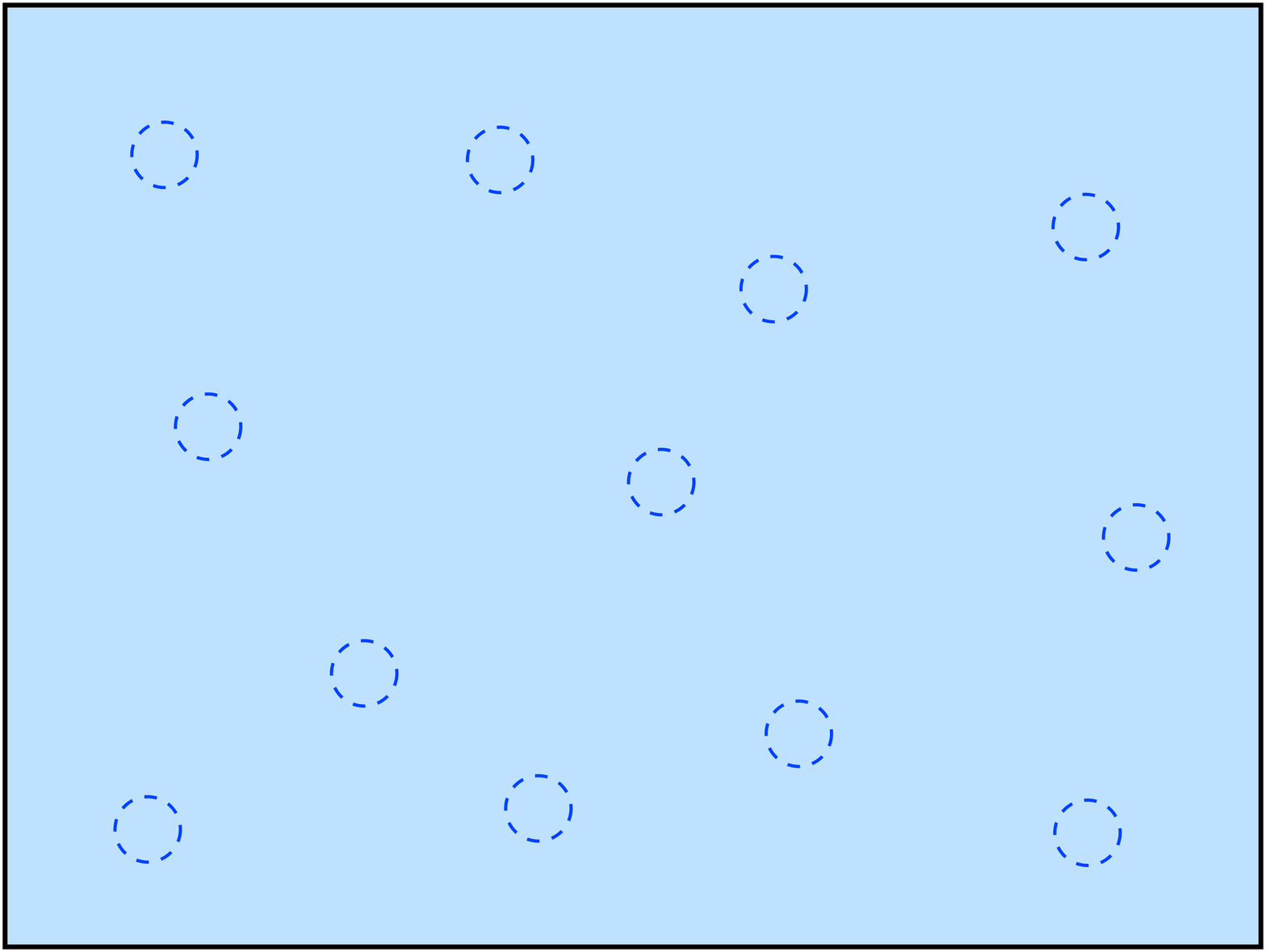}
\caption{For purposes of illustration, this schematic shows diffusion spreadability at different times for the special case
in which phase 2 is comprised of a spatial distribution of particles. The left panel depicts
the uniform concentration of the solute species within phase 2 (dark blue regions) at time $t=0$. The middle panel
depicts the spreading of diffusion information  at short times. The right panel depicts the uniform concentration of the solute
species throughout both phases (light blue region) in the infinite-time limit. The behavior of the spreadability 
${\cal S}(t)$ as a function of time is intimately related to the underlying microstructure. Section \ref{periodic} describes remarkable links between
 the spreadability ${\cal S}(t)$, covering problem of discrete geometry, and nuclear magnetic resonance (NMR) measurements.}
\label{cartoon}
\end{figure*}

Remarkably, the consequences of Prager's result are unknown because it has yet to be understood fundamentally or applied in any meaningful way.
The purpose of this investigation is to explore the fundamental theoretical and practical implications of 
the spreadability  ${\cal S}(t)$. We begin by generalizing Prager's formula (\ref{0}) to all space dimensions (Sec. \ref{direct}).
We then obtain a new Fourier representation
of the spreadability  ${\cal S}(t)$ (Sec. \ref{fourier}) in terms of the spectral density ${\tilde \chi_{_V}}({\bf k})$ (defined
in Sec. \ref{back}), which is obtainable from scattering experiments. There are many fundamental questions that we will explore. For example, 
what microstructural information is reflected by the spreadability  ${\cal S}(t)$?
What microstructures {\it maximize} spreadability  up to time $t$?
We determine microstructures for which the ``spreadability"
is ``fast" or ``slow," thereby gaining an understanding
of how the microstructure  affects such time-dependent diffusion processes.

Using the exact direct- and Fourier-space representations of the spreadability  (Sec. \ref{theory}),
 we  derive closed-form general asymptotic expansions of the spreadability  for any $d$ that apply at
short times and long times in terms of crucial small- and large-scale microstructural information,
respectively. We show that the small-time behavior of ${\cal S}(t)$ is determined by the derivatives of $S_2({\bf r})$
at the origin, the leading order term of order $t^{1/2}$ being proportional to the specific surface $s$ (interface area
per unit volume). By contrast, the corresponding long-time behavior is determined by the form of the spectral
density ${\tilde \chi_{_V}}({\bf k})$ at small wavenumbers.

We obtain exact results for ${\cal S}(t)$ for a variety of specific ordered and disordered model microstructures across dimensions that span
from hyperuniform to antihyperuniform media (Secs. \ref{app} and \ref{app-stealthy}).  Hyperuniform two-phase media are characterized by
an anomalous suppression of volume-fraction fluctuations relative to garden-variety disordered media \cite{Za09,To18a}
and can be endowed with novel physical properties \cite{To18a}; see Sec. \ref{back} for precise mathematical definitions.
For hyperuniform media, disordered or not, we show that the excess spreadability, ${\cal S}(\infty)-{\cal S}(t)$,
decays to its long-time behavior exponentially faster than that of any non-hyperuniform two-phase medium, the `slowest" being
antihyperuniform media (Sec. \ref{long}). The stealthy hyperuniform
class (see Sec. \ref{back}) is characterized by an excess spreadability with the fastest decay rate among all hyperuniform media and hence
all translationally invariant microstructures.  Specifically, ${\cal S}(t)$ for stealthy hyperuniform media
decays faster than any inverse power law (Sec. \ref{app-stealthy}), the latter of which applies to any {\it nonstealthy} disordered hyperuniform medium (Sec. \ref{long}). Thus, the spreadability provides a new dynamic-based figure
of merit to probe and classify the spectrum of possible microstructures  that span between hyperuniform and nonhyperuniform media.

We establish that the microstructures with ``fast" spreadabilities are also those that can be derived from efficient ``coverings" 
of Euclidean space $\mathbb{R}^d$ (Sec. \ref{covering}). 
Moreover, in  Sec. \ref{NMR}, we  identify a heretofore unnoticed fascinating connection between the spreadability  
${\cal S}(t)$ and noninvasive
nuclear magnetic resonance (NMR) relaxation measurements in physical and biological porous media \cite{Br79,Mit92b,Mi93,Se94,Or02,We05,No14}. 
We close with concluding remarks (Sec. \ref{discuss}), including a ``phase diagram" that schematically shows the spectrum
of spreadability regimes and its relationship to the spectrum of microstructures.

%In summary, our work reveals that the spreadability  across time scales is a new 
%and powerful figure of merit that enables one to probe dynamically salient microstructural information
%across length scales in complex two-phase media. This time-de

\section{Background}
\label{back}

\subsection{Correlation Functions}

A two-phase medium is fully statistically characterized by the $n$-point correlation functions \cite{To02a},
defined by
\begin{equation}
S_{n}^{(i)} \left( \mathbf{x}_1, ..., \mathbf{x}_n \right) \equiv
\left \langle {\cal I}^{(i)}(\mathbf{x}_1) \ldots  {\cal I}^{(i)}(\mathbf{x}_n) \right \rangle,
\end{equation}
where ${\cal I}^{(i)}({\bf x})$ is the {\it indicator function} for phase $i=1, 2$, $n=1,2,3,\ldots$,
and angular brackets denote an ensemble average.  The function $S_n^{(i)}({\bf x}_1, \ldots, {\bf x}_n )$ 
also has a probabilistic interpretation, namely, it is the
probability that the vertices of a polyhedron located
at ${\bf x}_1, \ldots, {\bf x}_n$ all lie in phase $i$.
For statistically homogeneous media, $S_{n}^{(i)} \left( \mathbf{x}_1, ..., \mathbf{x}_n \right)$
is translationally invariant and hence depends only on the relative displacements of the points.

The {\it autocovariance} function $\chi_{_V}({\bf r})$, which is directly related to the two-point function $S_2^{(i)}({\bf r})$
and plays a central role in this paper, is defined by
\begin{equation}
\chi_{_V}({\bf r}) \equiv S_2^{(1)}({\bf r})-\phi_1^2=S_2^{(2)}({\bf r})-\phi_2^2.
\label{covariance}
\end{equation}
Here, we have assumed statistical homogeneity.
At the extreme limits of its argument, $\chi_{_V}({\bf r})$ has the following asymptotic behavior:
$\chi_{_V}({\bf r}=0)=\phi_1\phi_2$ and  $\lim_{|{\bf r}| \rightarrow \infty} \chi_{_V}({\bf r})=0$ if
the medium possesses no long-range order.
If the medium is statistically homogeneous and isotropic, then the  autocovariance
function ${\chi_{_V}}({\bf r})$ depends only on the magnitude of its argument $r=|\bf r|$,
and hence is a radial function. In such instances, its slope at the origin is directly related
to the {\it specific surface} $s$, which is the interface area per unit volume. In particular, the well-known
three-dimensional asymptotic result \cite{De57} is easily obtained  in any space
dimension $d$:
\begin{equation}
\chi_{_V}({\bf r})= \phi_1\phi_2 - \kappa(d) s \;|{\bf r}| + {\cal O}(|{\bf r}|^2),
\label{specific}
\end{equation}
where
\begin{equation}
\kappa(d)= \frac{\Gamma(d/2)}{2\sqrt{\pi} \Gamma((d+1)/2)}.
\label{kappa}
\end{equation}

The nonnegative spectral density ${\tilde \chi}_{_V}({\bf k})$, which can be obtained from  scattering experiments \cite{De49,De57},
is  the Fourier transform of $\chi_{_V}({\bf r})$ at wave vector $\bf k$, i.e.,
\begin{equation}
{\tilde \chi}_{_V}({\bf k}) = \int_{\mathbb{R}^d} \chi_{_V}({\bf r}) e^{-i{\bf k \cdot r}} {\rm d} {\bf r} \ge 0, \qquad \mbox{for all} \; {\bf k}.
\label{spectral}
\end{equation}
 For isotropic media, the spectral density only depends
on the wavenumber $k=|{\bf k}|$ and, as a consequence of  (\ref{specific}), its decay in the large-$k$ limit is controlled
by the exact following power-law form:
\begin{equation}
{\tilde \chi}_{_V}({\bf k}) \sim \frac{\gamma(d)\,s}{k^{d+1}}, \qquad k \rightarrow \infty,
\label{decay}
\end{equation}
where
\begin{equation}
\gamma(d)=2^d\,\pi^{(d-1)/2} \,\Gamma((d+1)/2).
\end{equation}

\subsection{Hyperuniformity}

The hyperuniformity concept generalizes the traditional notion of long-range order in many-particle systems
to not only include all perfect crystals and perfect quasicrystals, but also exotic amorphous states of matter according to \cite{To03a,To18a}.
For two-phase heterogeneous media in $d$-dimensional Euclidean space $\mathbb{R}^d$, which include cellular solids, composites, and porous media, hyperuniformity is defined by the following infinite-wavelength  condition on the {\it spectral density} ${\tilde \chi}_{_V}({\bf k})$\cite{Za09, To18a}, i.e.,
\begin{equation}
\lim_{|{\bf k}|\to 0 }{\tilde \chi}_{_V}({\bf k}) = 0.
\label{condition}
\end{equation}
An equivalent definition of hyperuniformity is based on the local volume-fraction
variance $\sigma^2_{_V}(R)$  associated with a $d$-dimensional spherical observation window  of radius $R$.
A two-phase medium in $\mathbb{R}^d$ is hyperuniform if its variance
grows in the large-$R$ limit faster than $R^d$.
This behavior is to be   contrasted with those of  typical disordered two-phase media for which the variance decays  like
the inverse of the volume $v_1(R)$ of the spherical observation window, which is given by 
\begin{equation}
v_1(R)=\frac{\pi^{d/2} R^d}{\Gamma(1+d/2)}.
\label{v1}
\end{equation} 
The  hyperuniformity  condition  (\ref{condition})  dictates  that  the  direct-space 
autocovariance function $\chi_{_V}({\bf r})$  exhibits both positive  and  negative  correlations  such  that  its  
volume  integral  over all space is exactly zero \cite{To16b}, i.e.,
\begin{equation}
\int_{\mathbb{R}^d} \chi_{_V}({\bf r}) d{\bf r}=0,
\label{sum-rule}
\end{equation}
which  is a direct-space  sum  rule for hyperuniformity.

\subsection{Classification of Hyperuniform and Nonhyperuniform Media}
\label{class}

The hyperuniformity concept has led to a unified means to classify equilibrium and nonequilibrium states
of matter, whether hyperuniform or not, according to their large-scale fluctuation characteristics.
In the case of hyperuniform two-phase media  \cite{Za09,To18a},  there are three different scaling regimes (classes) that describe the associated large-$R$ behaviors of the volume-fraction variance when the spectral density goes to zero as a power-law scaling  ${\tilde \chi}_{_V}({\bf k})\sim |{\bf k}|^\alpha$ as $|\bf k|$ tends to zero:
\begin{align}  
\sigma^2_{_V}(R) \sim 
\begin{cases}
R^{-(d+1)}, \quad\quad\quad \alpha >1 \qquad &\text{(Class I)}\\
R^{-(d+1)} \ln R, \quad \alpha = 1 \qquad &\text{(Class II)}\\
R^{-(d+\alpha)}, \quad 0 < \alpha < 1\qquad  &\text{(Class III).}
\end{cases}
\label{eq:classes}
\end{align}
Classes I and III are the strongest and weakest forms of hyperuniformity, respectively.
Class I media include all crystal structures, many quasicrystal structures and exotic
disordered media \cite{Za09,To18a}. Stealthy hyperuniform  media are also of class I and are defined to be those that possess 
zero-scattering intensity for a set of wavevectors around the origin \cite{To16b}, i.e.,
\begin{equation}
{\tilde \chi}_{_V}({\bf k})=0 \qquad \mbox{for}\; 0 \le |{\bf k}| \le K.
\label{stealth}
\end{equation}
Examples of such media are periodic packings of spheres
as well as unusual disordered sphere packings derived from stealthy point patterns \cite{To16b,Zh16b}.

By contrast, for any nonhyperuniform two-phase system, it is straightforward to show,
using a similar analysis as for point configurations \cite{To21c}, that
the local variance has the following large-$R$ scaling behaviors:
\begin{align}  
\sigma^2_{_V}(R) \sim 
\begin{cases}
R^{-d}, & \alpha =0 \quad \text{(typical nonhyperuniform)}\\
R^{-(d+\alpha)}, & -d <\alpha < 0 \quad \text{(antihyperuniform)}.\\
\end{cases}
\label{sigma-nonhyper}
\end{align}
For a  ``typical" nonhyperuniform system, ${\tilde \chi}_{_V}(0)$ is bounded  \cite{To18a}. In {\it antihyperuniform} systems,
${\tilde \chi}_{_V}(0)$ is unbounded, i.e.,
\begin{equation}
\lim_{|{\bf k}| \to 0} {\tilde \chi}_{_V}({\bf k})=+\infty,
\label{antihyper}
\end{equation}
and hence  are diametrically opposite to hyperuniform systems.
Antihyperuniform systems include  systems at thermal critical points (e.g., liquid-vapor and magnetic critical points) \cite{St87b,Bi92}, fractals \cite{Ma82}, disordered non-fractals \cite{To21b},
and certain substitution tilings \cite{Og19}.

\section{Theory}
\label{theory}

\subsection{Generalization of Prager's formula for All  Dimensions}
\label{direct}

Using the $d$-dimensional Green's function for the time-dependent diffusion equation,
it is straightforward to generalize Prager's three-dimensional result for the spreadability ${\cal S}(t)$, given
by (\ref{0}), to any Euclidean space dimension $d$. After rearranging terms, we find that
\begin{equation}
{\cal S}(t)= \frac{1}{(4\pi D t)^{d/2}\, \phi_2} \int_{\mathbb{R}^d}[\phi_2- S_2({\bf r})] \exp[-r^2/(4Dt)] d{\bf r},
\label{1}
\end{equation}
where  it is to be noted that  ${\cal S}(\infty)=\phi_1$, i.e., the infinite-time value
of ${\cal S}(t)$. We note the identities
\begin{equation}
\frac{1}{(4\pi D t)^{d/2}} \int_{\mathbb{R}^d}  \exp[-r^2/(4Dt)] d{\bf r}=1 
\label{ident-1}
\end{equation}
and
\begin{equation}
\frac{1}{(4\pi D t)^{d/2}} \int_{\mathbb{R}^d}  r^2 \exp[-r^2/(4Dt)] d{\bf r}= 2\,d D\, t.
\label{ident-2}
\end{equation}
The second identity is nothing more than the mean-square displacement
of a freely diffusing particle  in the long-time limit.
Use of the first identity in (\ref{1}) yields the difference ${\cal S}(\infty)-{\cal S}(t)$, which we call the {\it excess spreadability},
to be given by
\begin{eqnarray}
{\cal S}(\infty)- {\cal S}(t)&=& \frac{1}{(4\pi D t)^{d/2}\, \phi_2} \int_{\mathbb{R}^d} \chi_{_V}({\bf r}) \exp[-r^2/(4Dt)] d{\bf r} \nonumber\\
&=&\frac{d \,\omega_d}{(4\pi Dt)^{d/2}\,\phi_2} \int_0^\infty r^{d-1} \chi_{_V}(r) \exp[-r^2/(4Dt)] dr, \nonumber\\
\label{2}
\end{eqnarray}
where
\begin{equation}
\omega_d =\frac{\pi^{d/2}}{\Gamma(1+d/2)}
\label{omega}
\end{equation}
is the volume of a $d$-dimensional sphere of unit radius and $\chi_{_V}({\bf r})$ is the autocovariance function, defined by (\ref{covariance}).
In the second line of (\ref{2}),
the autocovariance  $\chi_{_V}(r)$ is the radial function that depends on the distance $r \equiv |{\bf r}|$, which  results
from averaging the vector-dependent quantity  $\chi_{_V}({\bf r})$ over all angles, i.e.,
\begin{equation}
 \chi_{_V}(r)=\frac{1}{\Omega} \int_{\Omega}  \chi_{_V}({\bf r})\, d\Omega,
\end{equation}
where $d\Omega$ is the differential solid angle and
\begin{equation}
\Omega = \frac{d \pi^{d/2}}{\Gamma(1+d/2)}
\end{equation}
is the total solid angle contained in a $d$-dimensional sphere. It is important to stress that relation (\ref{2})
applies to all translationally invariant two-media, including periodic media.

\begin{figure}[bthp]
\includegraphics[  width=1.65in,clip=keepaspectratio]{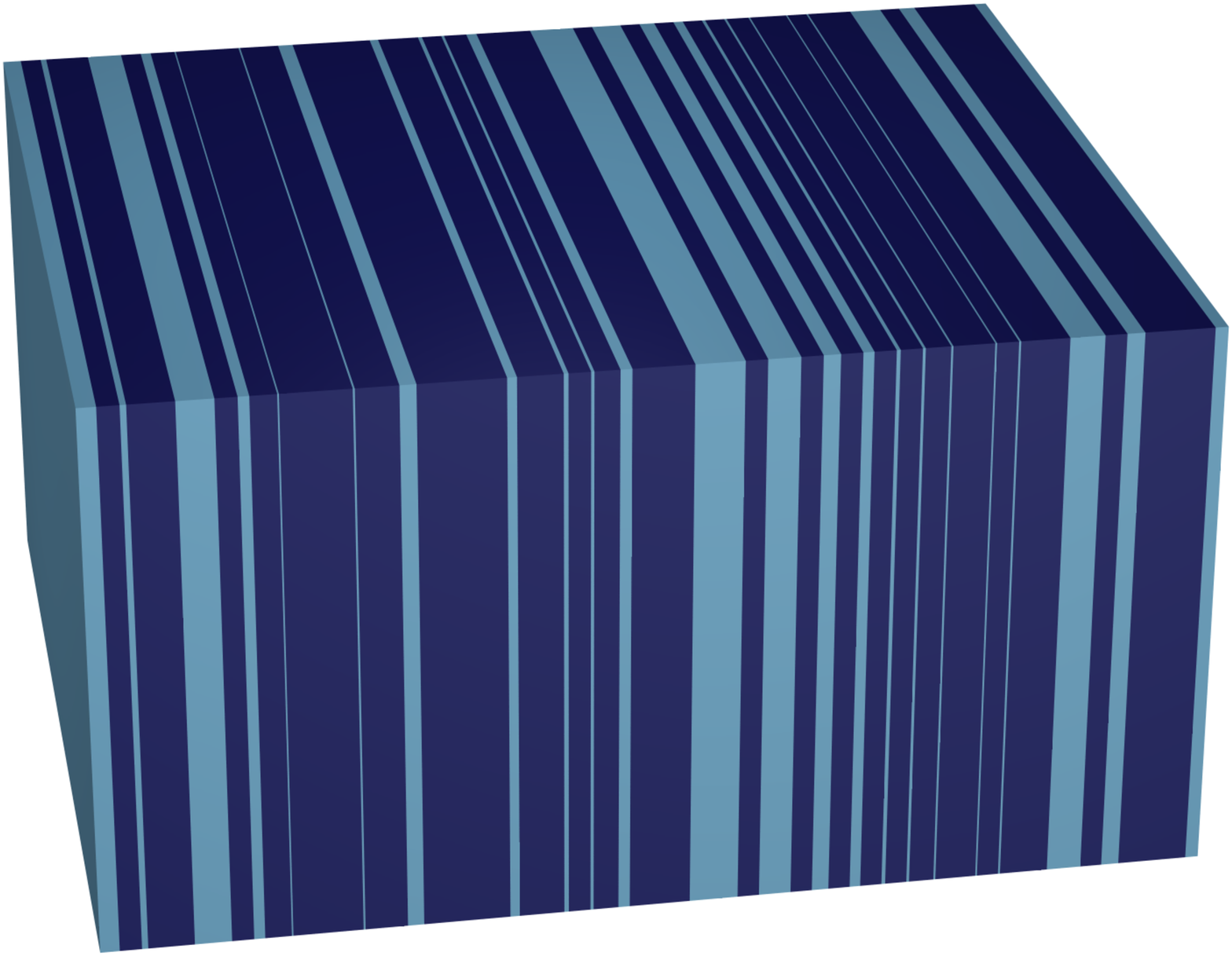}	\includegraphics[  width=1.65in,clip=keepaspectratio]{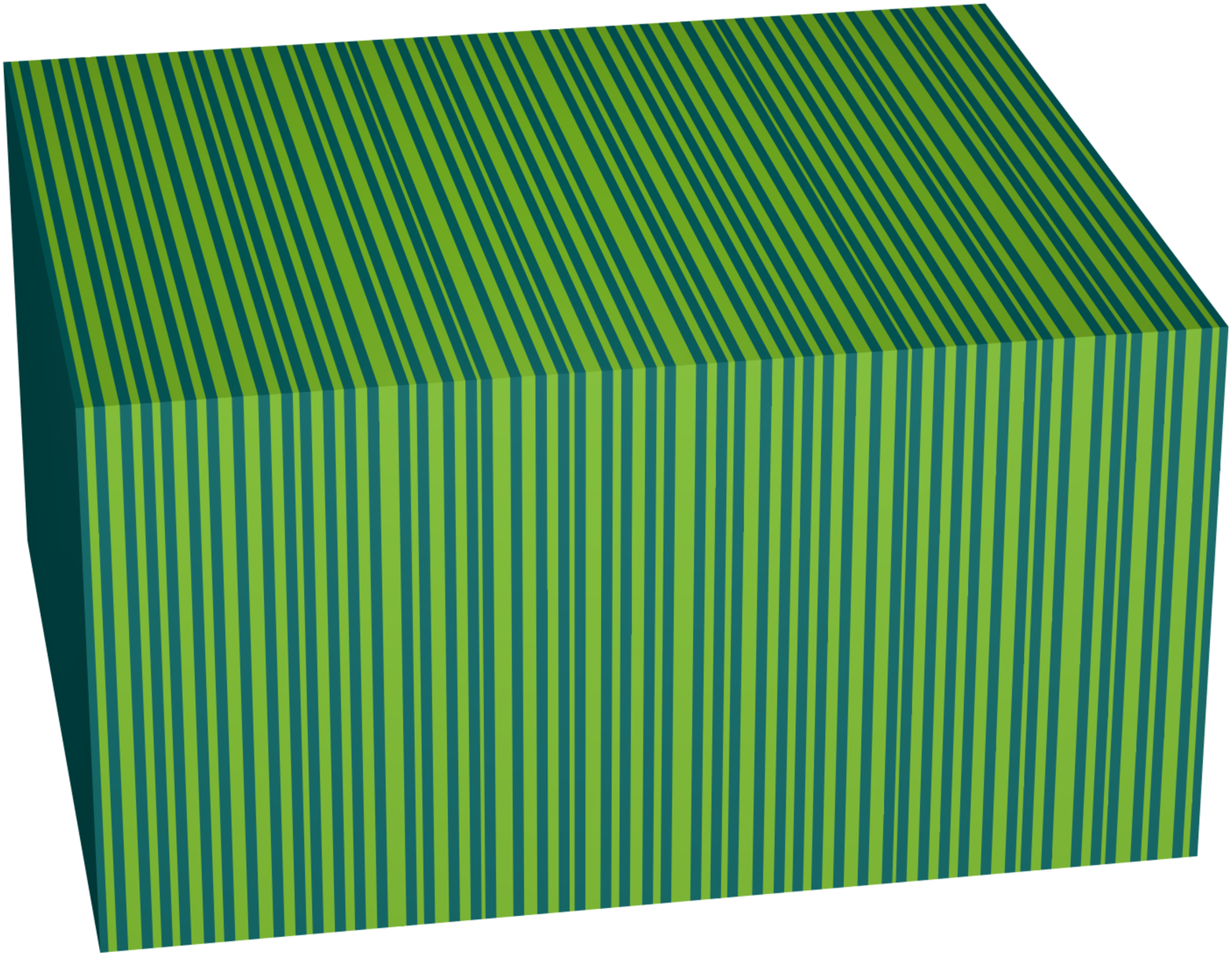}\\
\includegraphics[  width=1.65in,clip=keepaspectratio]{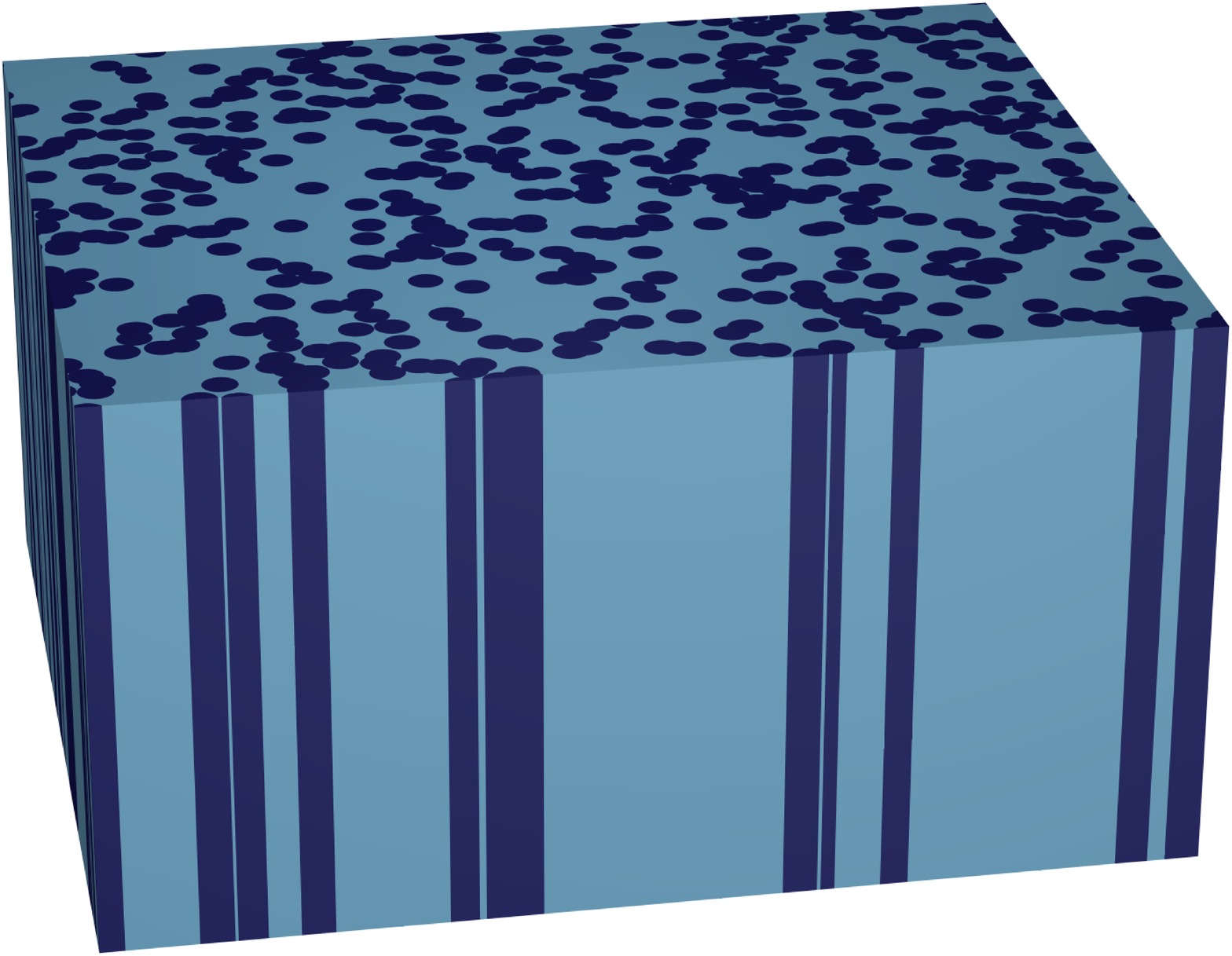}	\includegraphics[  width=1.65in,clip=keepaspectratio]{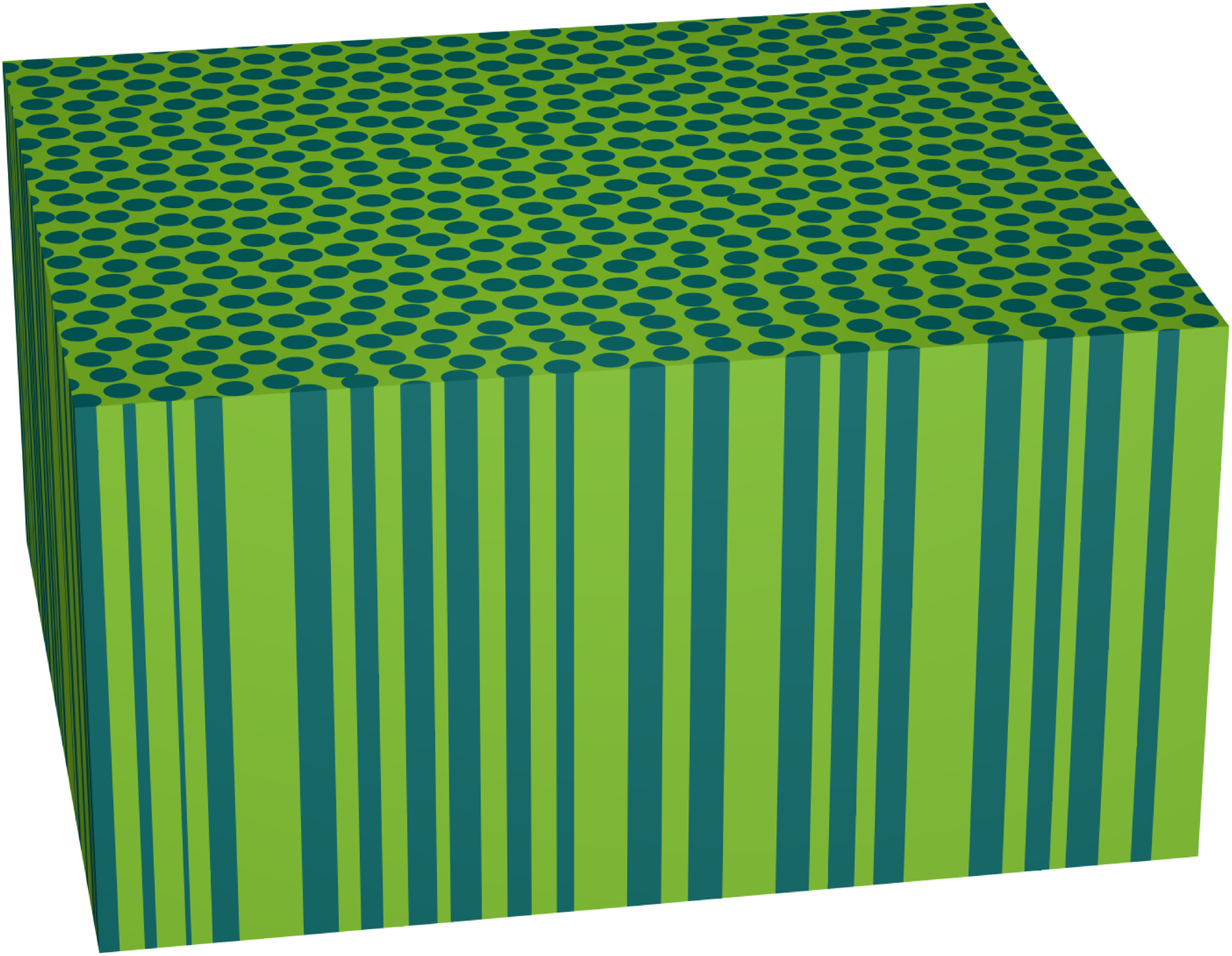}\\
\includegraphics[  width=1.65in,clip=keepaspectratio]{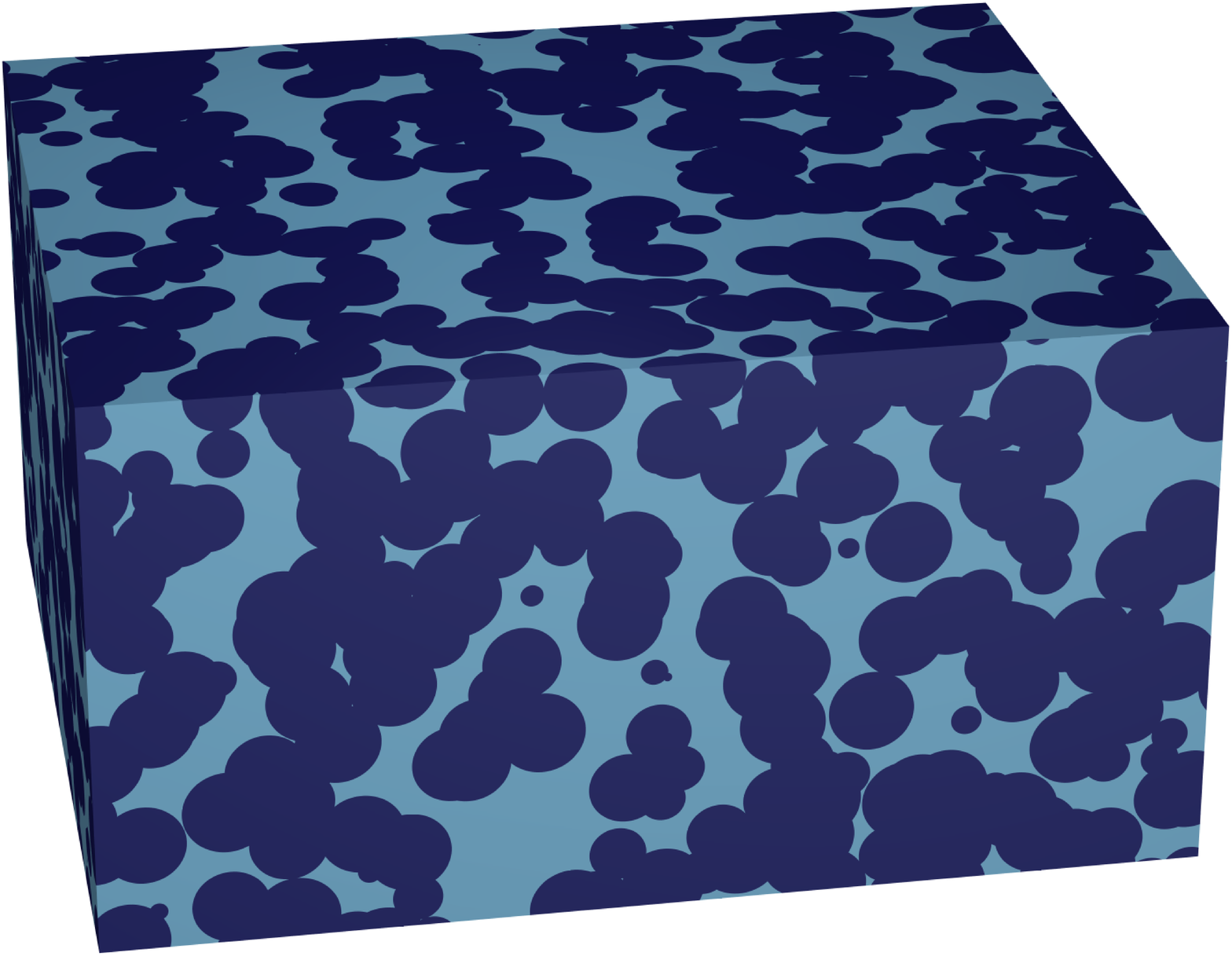}	\includegraphics[  width=1.65in,clip=keepaspectratio]{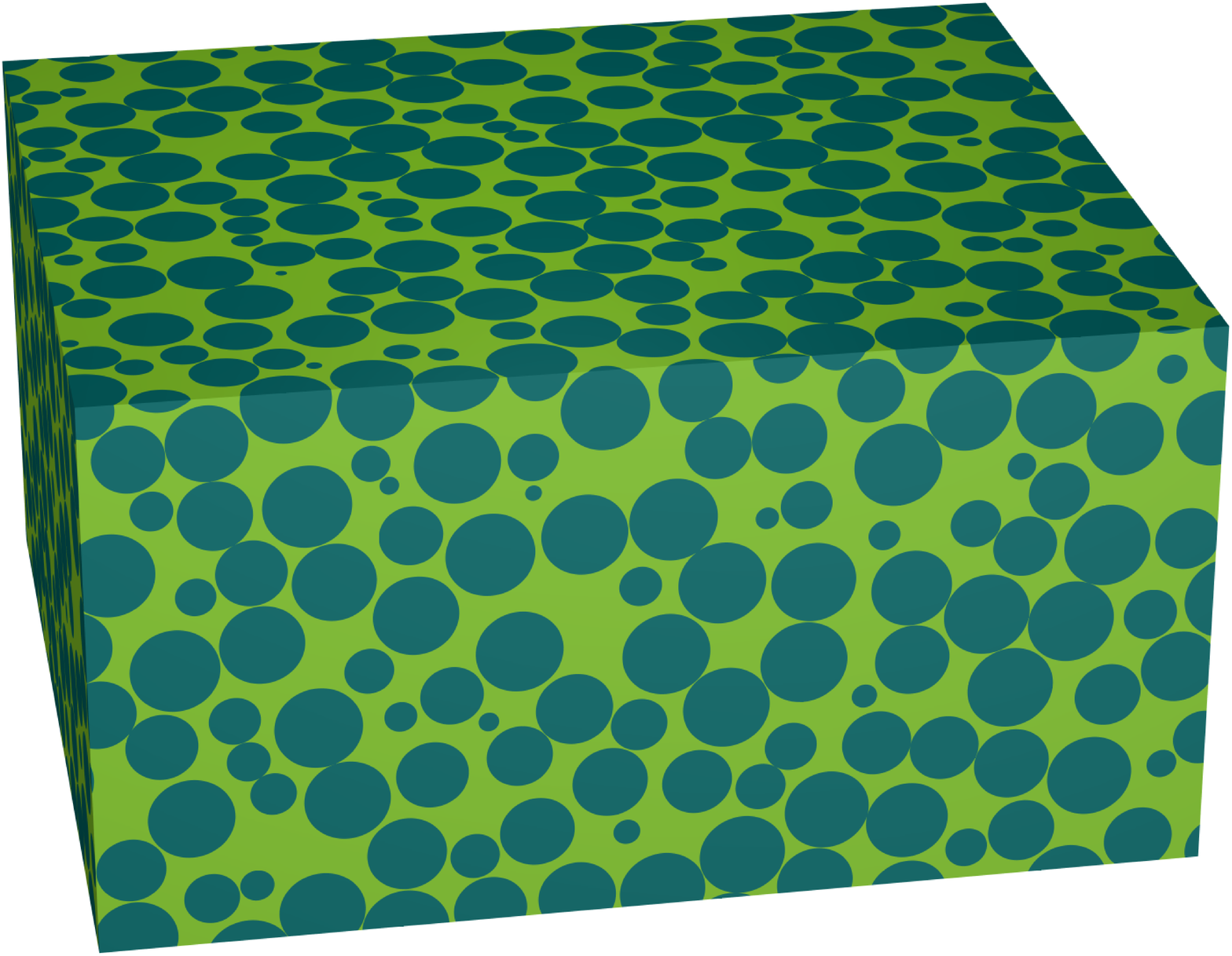}	
\caption{\footnotesize Models of nonhyperuniform and hyperuniform two-phase media with different symmetries are shown in 
blue (left) and green (right) colors, respectively. In each row, the spreadability is larger for the hyperuniform medium than that of the corresponding
nonhyperuniform medium, as proved in Sec. \ref{long}. Top row: 3D anisotropic stratified two-phase
media consisting of infinite parallel slabs of phases 1 and 2 ($\phi_2=0.5$) whose thicknesses are derived from nonhyperuniform overlapping rods \cite{To02a} (left) and hyperuniform
perturbed 1D integer lattice point patterns \cite{Kl20a} (right). Formulas (\ref{2}) and (\ref{3}) with $d=1$ for ${\cal S}(t)$ apply rigorously for these 3D anisotropic structures. Middle row:  3D anisotropic two-phase media
 ($\phi_2=0.5$)  with cylindrical symmetry obtained from nonhyperuniform
oriented overlapping circular cylinders \cite{To02a} (left) and stealthy and hyperuniform oriented nonoverlapping circular cylinders \cite{Zh16b,To21a} (right). Formulas (\ref{2}) and (\ref{3}) with $d=2$ for ${\cal S}(t)$ apply rigorously for these 3D anisotropic structures.
Bottom row: 3D isotropic two-phase media  ($\phi_2=0.636$) of overlapping spheres \cite{To02a} (left) and hyperuniform maximally random jammed spheres \cite{Kla14,To18b} (right).
}
\label{layered}
\end{figure}

Figure \ref{layered} shows examples of three-dimensional (3D) nonhyperuniform and hyperuniform media
with different symmetries for which formula (\ref{2}) for the spreadability rigorously applies.
It is noteworthy that the formula (\ref{2}),  as well as formula (\ref{3}) below, for one-dimensional (1D) cases (i.e., $d=1$) 
are also rigorously exact for the  idealized 
three-dimensional (3D) statistically {\it anisotropic} stratified two-phase
media of parallel slabs of phases 1 and 2, as illustrated in the top row of  Fig. \ref{layered}.
This fact is easily proved by employing the first line of formula (\ref{2}), for example, with $d=3$
using Cartesian coordinates, and then recognizing that  the vector-dependent quantity $\chi_{_V}({\bf r})$
is independent of the components of $\bf r$ in the directions orthogonal to the slab normal.
Similarly, formulas  (\ref{2}) and (\ref{3}) for two-dimensional (2D) cases (i.e., $d=2$)
are rigorously exact for the  idealized  three-dimensional (3D) {\it anisotropic} media 
  that possess transverse isotropy with respect to  an axis of symmetry, as illustrated
in the middle row of  Fig. \ref{layered}. The bottom row of Fig. \ref{layered}
shows examples of 3D statistically isotropic disordered nonhyperuniform and hyperuniform media.

%When the medium is statistically homogeneous and isotropic media, the exact relation (\ref{1})
%simplifies as follows:
%\begin{eqnarray}
%{\cal S}(\infty)- {\cal S}(t) &=&\frac{d \,\omega_d}{(4\pi Dt)^{d/2}\,\phi_2} \int_0^\infty r^{d-1} \chi_{_V}(r) \exp[-r^2/(4Dt)] dr,
%\label{3}
%\end{eqnarray}
%where 
%, and $r \equiv |\bf r|$.

\subsection{Fourier Representation of the Spreadability}
\label{fourier}

Here, we obtain a Fourier representation of the spreadability, which is useful when scattering information
is available.
By Parseval's theorem, the direct-space relation (\ref{2}) for the spreadability can be re-expressed in Fourier space as
\begin{eqnarray}
{\cal S}(\infty)- {\cal S}(t)&=&\frac{1}{(2\pi)^d\,\phi_2} \int_{\mathbb{R}^d} {\tilde \chi}_{_V}({\bf k}) \exp[-k^2 Dt] d{\bf k} \nonumber\\
&=& \frac{d\,\omega_d}{(2\pi)^d\,\phi_2} \int_0^\infty k^{d-1} {\tilde \chi}(k) \exp[-k^2 Dt] dk, \nonumber\\
\label{3}
\end{eqnarray}
where ${\tilde \chi}_{_V}({\bf k})$ is the spectral density, which is the Fourier transform of $\chi_{_V}({\bf r})$,
and $\bf k$ is the wave vector. In the second line of (\ref{3}),
the spectral density ${\tilde \chi}_{_V}(k)$ is the radial function that depends on the wavenumber $k \equiv |{\bf k}|$, which  results
from averaging the vector-dependent quantity ${\tilde \chi}_{_V}({\bf k})$ over all angles, i.e.,
\begin{equation}
{\tilde \chi}_{_V}(k)=\frac{1}{\Omega} \int_{\Omega}  {\tilde \chi}_{_V}({\bf k}) \, d\Omega,
\end{equation}
$d\Omega$ is the differential solid angle. Now, since ${\tilde \chi}_{_V}({\bf k})$ is nonnegative for all $\bf k$,
 the integrand of (\ref{3}) is nonnegative and decreases with increasing $t$. Thus,  the 
 excess spreadability is a monotonically decreasing function of time and is itself nonnegative, i.e.,
\begin{equation}
{\cal S}(\infty) -{\cal S}(t) \ge 0 \qquad \mbox{for all}\; t
\label{pos}
\end{equation}
or, equivalently,
\begin{equation}
{\cal S}(t) \le \phi_1 \qquad \mbox{for all}\; t.
\end{equation}
In summary, we can ascertain the spreadability
exactly for any microstructure across spatial dimensions using knowledge of the
corresponding autocovariance via relation (\ref{2}) or the spectral density via (\ref{3}).

%When the two-phase medium is statistically homogeneous and isotropic, relation (\ref{4}) simplifies as
%\begin{eqnarray}
%{\cal S}(\infty)- {\cal S}(t)&=& \frac{d\,\omega_d}{(2\pi)^d\,\phi_2} \int_0^\infty k^{d-1} {\tilde \chi}_{_V}(k) \exp[-k^2 Dt] dk,
%\label{5}
%\end{eqnarray}
%where $k \equiv |\bf k|$ is the wavenumber.

%\subsection{Generalization of Prager's expression for ${\cal S}(t)$  to any $d$ and its Fourier representation}

\subsection{Small-Scale Structure via Short-Time Behavior of ${\cal S}(t)$}
\label{short}

To obtain the short-time asymptotic behavior of ${\cal S}(t)$ for statistically homogeneous media, 
we  recognize that the Gaussian term $\exp[-r^2/(4 D t)]$ in the direct-space representation
of the spreadability (\ref{2}) is nonnegligibly small for short times for distances only near the spatial
origin ($r=0$). Therefore, the short-time behavior of the integral in (\ref{2}) is determined by the small-$r$ expansion of $\chi_{_V}(r)$ about $r=0$:
\begin{equation}
\chi_{_V}(r) = \phi_1\phi_2 - \frac{\omega_{d-1}}{\omega_d \,d} \,s\, r + \sum_{n=2}^N s_n r^n,
\label{s}
\end{equation}
where  $s$ is the specific surface and the  coefficient $s_n=(d^n \chi_{_V}(r)/dr^n)_{r=0}$ is the $n$th order derivative
at the origin. 
%This small-$r$ behavior implies that the   spectral density 
%at very large wavenumbers for disordered media is dominated by the following power-law behavior:
%\begin{equation}
%{\tilde \chi}_{_V}(k) \sim \frac{2^d\, \pi^{d-1}\,s }{ d\,\omega_d\, k^{d+1}} \quad (k \to \infty).
%\end{equation}
Substitution of (\ref{s}) into (\ref{2}) yields the following exact
asymptotic expansion of ${\cal S}(t)$ for any $d$:
\begin{widetext}
\begin{equation}
{\cal S}(t) = \frac{s}{\phi_2} \left(\frac{Dt}{\pi}\right)^{1/2} + \frac{d\, \omega_d}{\pi^{d/2} \phi_2} \sum_{n=2}^{N} 2^{n-1}\, s_{n}\,\Gamma((n+d)/2)\, (Dt)^{n/2} \quad (t \to 0),
\label{small}
\end{equation}
\end{widetext}
where we have employed the integral identity
\begin{eqnarray}
\frac{1}{(4\pi Dt)^{d/2}} \hspace{-0.2in}&&\int_{0}^\infty r^m \exp[-r^2/(4Dt)] dr  \nonumber\\
&=&2^{m-d}\, {\pi^{-d/2} \Gamma((m+1)/2)}\, (Dt)^{(m+1-d)/2}, \nonumber \\
\end{eqnarray}
and $m$ is a nonnegative integer. It is noteworthy that if the upper limit  $N$ in the sum (\ref{small})
is infinite, i.e., the $s_n$ exist for all $n \ge 2$, formula (\ref{small}) is an exact convergent series
representation of the spreadability for all times.  
The first two terms of the short-time asymptotic expansion (\ref{small}) are explicitly given by
\begin{equation}
{\cal S}(t) = \frac{s}{\phi_2} \left(\frac{Dt}{\pi}\right)^{1/2} - \frac{2 \,d\,s_2}{\phi_2}\, (D t) + {\cal O}(Dt/a^2)^{3/2},
\label{short-time}
\end{equation}
where $a$ is some characteristic heterogeneity length scale. Note that
the leading term is of order $t^{1/2}$, independent of the space dimension,
and proportional to the specific surface $s$, which is intuitively clear, since the solute species
is only just emerging from phase 2 in the immediate vicinity of the two-phase interface.   The term of order $t$ is determined by the curvature of $\chi_{_V}(r)$
at the origin due to the presence of the coefficient $s_2$. 
%Furthermore, the term of order
%$t$ is directly proportional to the mean-square displacement of the diffusing particle.

%where
%\begin{equation}
%m_{d-1}= \frac{1}{d \, \omega_d}.
%\end{equation}

\subsection{Large-Scale Structure via Long-Time Behavior of ${\cal S}(t)$}
\label{long}

The long-time behavior of the spreadability ${\cal S}(t)$ is determined by the large-scale structural
characteristics of the two-phase medium. Specifically, 
we see that the integrand of the Fourier representation  (\ref{3}) of the spreadability is nonnegligibly small at long times for  wavenumbers 
in the vicinity of the origin, i.e., the behavior of the spectral density ${\tilde \chi}_{_V}(k)$ in the
infinite-wavelength limit.   In the special
situation in which ${\tilde \chi}_{_V}(k)$ is an analytic function at the origin, the spectral density admits a Taylor series
expansion in only even powers of $k$ and whose coefficients depend on certain moments of the autocovariance function $\chi_{_V}(r)$,
all of which must exist. 
Specifically, using (\ref{3}), we find the following exact series representation of the excess spreadability ${\cal S}(\infty)-{\cal S}(t)$:
%Substitution of the Taylor series of $\exp[-r^2/(4\pi D t]$ in  (\ref{3}) yields the following exact series
%representation of ${\cal S}(t)$:
\begin{eqnarray}
{\cal S}(\infty)- {\cal S}(t)
% &=&\frac{d \,\omega_d}{(4\pi Dt)^{d/2}\,\phi_2} \int_0^\infty r^{d-1} \chi_{_V}(r) \sum_{n=0}^\infty \frac{(-1)^n r^{2n}}{n!(4 D t)^n} \nonumber \\
&=& \frac{d \,\omega_d}{(4\pi Dt)^{d/2}\,\phi_2}\sum_{n=0}^\infty \frac{(-1)^n M_{2n+d-1}(\chi_{_V})}{n!(4 D t)^n} ,
\label{6}
\end{eqnarray}
where 
\begin{equation}
M_n(\chi_{_V})= \int_0^\infty r^n \chi_{_V}(r) dr
\end{equation}
is the $n$th moment of $\chi_{_V}(r)$.
% which must exist for any $n$ for this analysis to be valid. 
Observe now that truncation of the infinite series (\ref{6}) yields the long-time asymptotic expansion
of the excess spreadability. The first few terms of this asymptotic expansion are explicitly given by
\begin{widetext}
\begin{eqnarray}
{\cal S}(\infty)- {\cal S}(t)&=&\frac{d \,\omega_d}{(4\pi Dt)^{d/2}\,\phi_2} \left[M_{d-1}(\chi_{_V}) - \frac{M_{d+1} (\chi_{_V})}{4 D t} +\frac{M_{d+3}(\chi_{_V})}{32 (Dt)^2} - \cdots\right] \quad (t\to \infty).
\label{7}
\end{eqnarray}
\end{widetext}
Note that since the moment $M_{d-1}(\chi_{_V})$ is nonnegative, then  the leading-order term of the sum is of order $t^{-d/2}$
whenever the system is nonhyperuniform, i.e., $M_{d-1}(\chi_{_V})$ does not vanish, and all moments exist.

Now we recognize that if this type of two-phase media is hyperuniform, then $M_{d-1}(\chi_{_V})$ in (\ref{7}) vanishes, implying 
that the leading-order term of the sum that involves the moment $M_{d+1}(\chi_{_V}) $ is of order $t^{-(d+2)/2}$, i.e.,
\begin{widetext}
\begin{eqnarray}
{\cal S}(\infty)- {\cal S}(t)&=& \frac{d \, \omega_d}{4  (4 \pi)^{d/2}  (Dt)^{d/2+1}\,\phi_2} \left[- M_{d+1}(\chi_{_V})
+\frac{M_{d+3}(\chi_{_V})}{8Dt} + \cdots\right] \quad (t\to \infty).
\label{hyp}
\end{eqnarray}
\end{widetext}
In light of the nonnegativity condition (\ref{pos}), the moment $M_{d+1}(\chi_{_V}) $ must be negative for a hyperuniform medium.
Moreover, since the spectral density ${\tilde \chi}_{_V}(k)$ is analytic at $k=0$ [i.e, all moments of $\chi_{_V}(r)$ exist], then it follows that
${\tilde \chi}_{_V}(k) \propto - M_{d+1}(\chi_{_V}) k^2$ in the limit $k \to 0$, and hence the two-phase medium  
is  hyperuniform of class I. Thus, we see that for such hyperuniform media, disordered or not, ${\cal S}(\infty)- {\cal S}(t)$
decays to its long-time behavior exponentially faster than that of any non-hyperuniform two-phase medium.
%Among this restricted class of hyperuniform media, the one that minimizes  $|M_{d+1}(\chi_{_V})|$, appropriately normalized, for some $\phi_2$ is the optimal solution.
%This is an open question. 

Now we consider the more general class of two-phase media in which the spectral density may be a nonanalytic function
at the origin such that it obeys the following power-law scaling in the infinite-wavelength limit: 
\begin{equation}
\lim_{|{\bf k}| \to {\bf 0}} {\tilde \chi}_{_V}({\bf k}) = B |{\bf k} a|^{\alpha},
\label{power}
\end{equation}
where $B$ is a positive dimensionless constant, $\alpha$ is an exponent that lies in the interval $(-d,\infty)$,
and $a$ represents some characteristic heterogeneity length scale. Antihyperuniform
media constitute cases in which $-d  < \alpha <0$. The case $\alpha=0$
corresponds to nonhyperuniform media, while the cases $\alpha >0$ 
correspond to hyperuniform media that may belong to class I, II or 	III (see Sec. \ref{class}).
  This small-wavenumber  behavior enables us to determine the more general long-time asymptotic behavior
of ${\cal S}(t)$ using the Fourier representation (\ref{3}). Specifically, we  find the
following general asymptotic expansion:
\begin{widetext}
\begin{equation}
{\cal S}(\infty)- {\cal S}(t) =  \frac{ B\, \Gamma((d+\alpha)/2)\,\phi_2}{2^d\,{\pi}^{d/2}\,\Gamma(d/2)\, (Dt/a^2)^{(d+\alpha)/2}} 
+{o}\left((Dt/a^2)^{-(d+\alpha)/2}\right) \quad (Dt/a^2 \gg 1),
\label{long-time}
\end{equation}
\end{widetext}
where ${o}(x)$ signifies all terms of order less than $x$.
Thus, we see that the long-time asymptotic behavior of ${\cal S}(t)$ is determined by the exponent $\alpha$
and the space dimension $d$, i.e., at long times, ${\cal S}(t)$ approaches the value $\phi_1$ with a power-law decay 
$1/t^{(d+\alpha)/2}$, implying a faster decay as $\alpha$ increases for some dimension $d$.
When $\alpha$ is bounded and positive, this result means that class I hyperuniform media
has the fastest decay, followed by class II and then class III, which has the slowest decay
among hyperuniform media. Of course, antihyperuniform media with $\alpha \to -d$ 
has the slowest decay among all translationally invariant media.
In the {\it  stealthy limit} in which $\alpha \to \infty$, the predicted infinitely-fast
inverse-power decay rate implies that the infinite-time aysmptote is approached exponentially
fast. This result will be demonstrated explicitly in the case of periodic  media, which are stealthy, as well
as disordered stealthy hyperuniform media.

\section{Applications to Nonhyperuniform, Hyperuniform and Antihyperuniform Media}
\label{app}

\subsection{Standard Nonhyperuniform Media}
\label{Debye}

\begin{figure}[t]
\centerline{\includegraphics[width=3.4in,keepaspectratio,clip=]{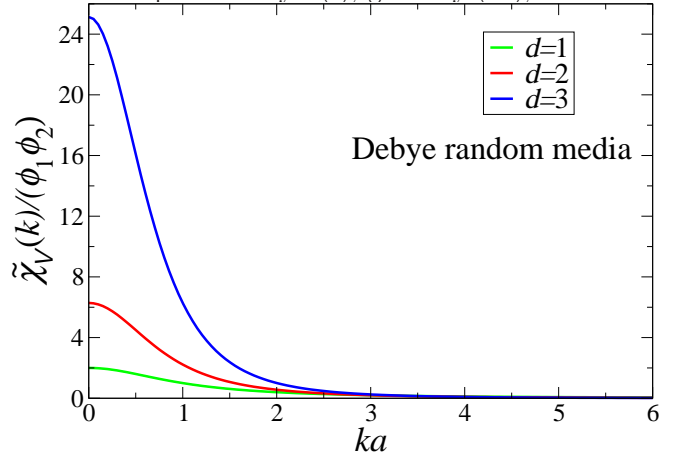}}
\caption{ The scaled spectral density ${\tilde \chi}_{_V}(k)/(\phi_1\phi_2)$ versus wavenumber $k$ for Debye random media for the
first three space dimensions, as obtained from (\ref{debye-spectral}). 
%Here we set the length scale $a$ in (\ref{debye}) to be unity. 
}
\label{debye-spec}
\end{figure}

It is instructive to first consider the spreadability ${\cal S}(t)$ for
 models of typical nonhyperuniform two-phase media. Prototypical
examples are  Debye random media \cite{Ye98a}, which are defined
entirely by the following monotonic radial autocovariance function:
\begin{equation}
\chi_{_V}(r) =\phi_1\phi_2 \exp(-r/a).
\label{debye}
\end{equation}
Such media can never be hyperuniform because  the sum rule (\ref{sum-rule}) requires
both positive and negative correlations \cite{To16b}. Debye et al. \cite{De57}
hypothesized the simple exponential form (\ref{debye})  to model three-dimensional media with
phases of ``fully random shape, size, and distribution.” It was many years
after their 1957 study that such autocovariance functions were shown to be realizable
in two \cite{Ye98a,Chi13,Ma20b} and three \cite{Ji07,To20a} dimensions.
The corresponding spectral density is given by
\begin{equation}
{\tilde \chi}_{_V}(k) = \frac{ \phi_1\phi_2\, 2^d \,\pi^{d-1}\, a^d}{\omega_{d-1}\,[1+(ka)^2]^{(d+1)/2}}.
\label{debye-spectral}
\end{equation}
Therefore, for small wavenumbers, 
\begin{equation}
{\tilde \chi}_{_V}(k) = \phi_1\phi_2 \frac{2^d \pi^{d-1}\,a^d}{\omega_{d-1}}   [1 -\frac{(d+1)}{2} (ka)^2 + {\cal O}((ka)^4)]
\end{equation}
so that ${\tilde \chi}_{_V}(0)=\phi_1\phi_2 2^d \pi^{d-1}\,a^d/\omega_{d-1}$.
The spectral density is plotted in Fig. \ref{debye-spec} for the first three space dimensions. 
We observe that Debye random media  departs from hyperuniformity superexponentially fast as the space dimension increases;
specifically, ${\tilde \chi}_{_V}(0)/(\phi_1\phi_2a^d) \sim  \sqrt{2} [2\pi d/\exp(1)]^{d/2}$ for large $d$.
%and
%\begin{equation}
%{\tilde \chi}_{_V}(k) \sim \frac{\phi_1\phi_2\, c_d}{ a k^{d+1}}=  \frac{2^d\, \pi^{d-1}\,s }{ d\,\omega_d\, k^{d+1}} \quad (k \to \infty).
%\end{equation}

It is convenient to rewrite the direct-space representation of the spreadability ${\cal S}(t)$, given by
(\ref{2}), as follows:
\begin{equation}
{\cal S}(\infty) -{\cal S}(t) =\frac{d\omega_d \phi_1}{(4\pi D t/a^2)^{d/2}} I_d(t),
\end{equation}
where
\begin{equation}
I_d(t) =\frac{1}{a^d}\int_0^\infty r^{d-1} \exp(-r/a) \exp[-r^2/(4Dt)] dr.
\end{equation}
We can obtain a closed-form exact expression for ${\cal S}(t)$ for Debye random media for any $d$ using the recurrence relation
\begin{equation}
I_{d+2}(t)=  \frac{2Dt}{a^2}\left[ d\, I_{d} - I_{d+1} \right].
\label{recur}
\end{equation}
Specifically, the explicit expressions
\begin{equation}
I_1(t)=\exp(Dt/a^2)\,\sqrt{\pi Dt/a^2}\,\left[1 - \mbox{erf}(\sqrt{Dt/a^2})\right]
\end{equation}
and
\begin{equation}
I_2(t)= \frac{2Dt}{a^2}\left\{1-\exp(Dt/a^2)\,\sqrt{\pi Dt/a^2}\,\left[1- \mbox{erf}(\sqrt{Dt/a^2})\right]\right\},
\end{equation}
for the first two dimensions combined with the recurrence relation (\ref{recur}) enables one to obtain $I_d$ for any $d \ge 3$. For example, for $d=3$, we have
\begin{widetext}
\begin{equation}
I_3(t)= \frac{2Dt}{a^2}\Big\{ \exp(Dt/a^2)\,\sqrt{\pi Dt/a^2}\,\left[1 -\mbox{erf}(\sqrt{Dt/a^2})\right]\left[1+2Dt/a^2\right]- 2Dt/a^2\Big\}.
\end{equation}
\end{widetext}

We also note that the $n$th moment of the autocovariance of Debye random media for any $d$ is given by
\begin{equation}
M_n(\chi_{_V})= \phi_1 \phi_2\, n! \,a^{n+1}.
\label{debye-mom}
\end{equation}
This result enables us to obtain another exact representation of the spreadability
via the infinite series (\ref{6}).

\begin{figure}[bthp]
\centerline{\includegraphics[width=3.4in,keepaspectratio,clip=]{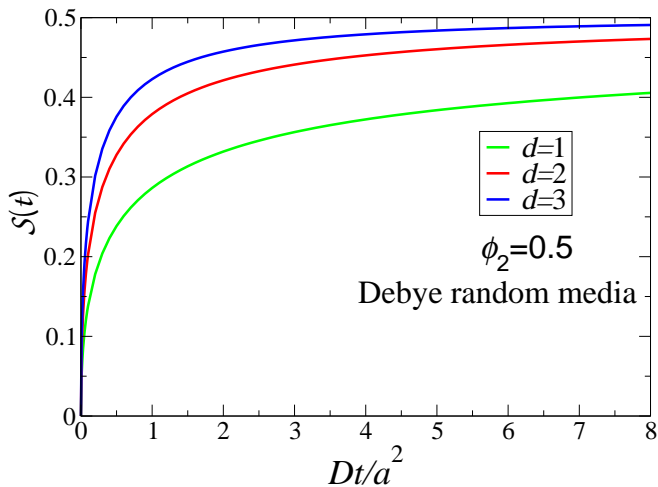}}
\caption{ The spreadability ${\cal S}(t)$ versus dimensionless time $D t/a^2$ for small to intermediate times for Debye random media in the
first three space dimensions. }
\label{debye-spread}
\end{figure}

For any space dimension $d$, the short-time behavior of the ${\cal S}(t)$ is given by
\begin{equation}
{\cal S}(t) = \frac{s}{\phi_2} \left(\frac{Dt/a^2}{\pi}\right)^{1/2} -\frac{d}{\phi_2} \left(\frac{D t}{a^2}\right) + {\cal O}((Dt/a^2)^{3/2}),
\end{equation}
where
\begin{equation}
s=  \frac{\phi_1\phi_2 \omega_d d}{\omega_{d-1} \, a}
\end{equation}
is the specific surface for a Debye random medium  and we have used (\ref{short-time}).
Employing (\ref{6}) and (\ref{debye-mom}), we see that the first two terms of the  long-time asymptotic expansion of the spreadability are given by
\begin{widetext}
\begin{equation}
{\cal S}(\infty)-{\cal S}(t) = \frac{(d-1)!\,d \omega_d \phi_2}{(4\pi Dt/a^2)^{d/2}} -\frac{(d+1)!\,d \omega_d\phi_2}{(4\pi Dt/a^2)^{(d+2)/2}} +{\cal O}\left((Dt/a^2)^{-(d+4)/2}\right).
\end{equation}
\end{widetext}
Figure \ref{debye-spread} shows the small- and intermediate-time behaviors
of the spreadability for Debye random media in the first three space dimensions.
It is seen that the effect of increasing dimensionality is to increase the spreadability
for a fixed  time for almost all times, namely, for dimensionless times $Dt/a^2> 1$.

\subsection{Disordered Hyperuniform Media}
\label{disorder-hyper}

To model hyperuniform two-phase media in $\mathbb{R}^d$, Torquato \cite{To16b} considered the following  family of autocovariance
functions:
\begin{equation}
 \frac{\chi_{_V}(r)}{\phi_1\phi_2}=c \,  e^{-r/a}\cos(qr +\theta),
\label{auto-hyper}
\end{equation}
where the parameters  $q$ and $\theta$ are the wavenumber and phase associated with the oscillations of $\chi_{_V}(r)$, respectively,
$a$ is a correlation length and $c$ is a normalization constant to be chosen so that the right-hand side
of (\ref{auto-hyper}) is unity for $r=0$.  In the special case in which $\theta=0$ and $c=1$,
Torquato showed that the corresponding autocovariance function 
satisfies all of the necessary realizability conditions and hyperuniformity constraint (\ref{sum-rule}) for $d =2$
if $(qa)^2=1$ and for $d=3$ if $(qa)^2=1/3$. Thus, the spectral densities for $d=2$ and $d=3$ are respectively given by
\begin{equation}
\frac{{\tilde \chi}_{_V}(k)}{\phi_1\phi_2}= \frac{ 2\pi (ka)^2 [A(k)+B(k)] + 4\pi [A(k)-B(k)]\, a^2}{[(ka)^4+4] [A^2(k)+B^2(k)] },
\label{spec2-hyper}
\end{equation}
and
\begin{equation}
\frac{{\tilde \chi}_{_V}(k)}{\phi_1\phi_2}= \frac{216\pi \, [3 (ka)^2+8](ka)^2\,a^3}{81 (ka)^8+216 (ka)^6+ 432 (ka)^4+ 384 (ka)^2+256},
\label{spec3-hyper}
\end{equation}
where
\begin{equation}
A(k)=\sqrt{(ka)^2/2+ \sqrt{(ka)^4+4}/2}, \qquad B(k)=A^{-1}(k).
\end{equation}

It was shown that for the special case $\theta=0$ and $d=1$, the function (\ref{auto-hyper})
does not satisfy the hyperuniformity constraint for any values of the parameters
$q$ and $\theta$. However, we note here that (\ref{auto-hyper}) meets all of the known realizability conditions 
and the hyperuniformity constraint for $d=1$, provided that the phase is given by
$\theta =\tan^{-1}\left(1/(qa)\right)$, implying that the normalization constant is
$c= [1+ (qa)^2]^{1/2}/(qa)$.  For concreteness, we set $qa=1$, and hence $c=2$ and $\theta=\pi/4$.
Taking the Fourier transform of (\ref{auto-hyper}) with these parameters  yields
the spectral density to be given by 
\begin{equation}
\frac{{\tilde \chi}_{_V}(k)}{\phi_1\phi_2}=  \frac{4\, (ka)^2\, a}{(ka)^4 +4}.
\label{spec1-hyper}
\end{equation}
Substitution of this expression into (\ref{3}) yields the following exact
formula for the spreadability 
\begin{equation}
{\cal S}(\infty)-{\cal S}(t)= \frac{4 \phi_1 \,\sqrt{t}}{\sqrt{\pi}} [s_{1,1/2}(2t)-1],
\end{equation}
where $s_{\mu,\nu}(x)$ is the Lommel function of the second kind \cite{Ab72}.

\begin{figure}[bthp]
\centerline{\includegraphics[width=3.6in,keepaspectratio,clip=]{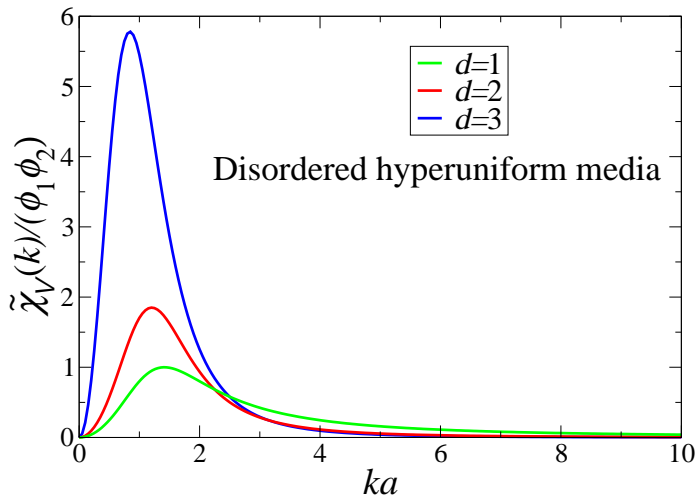}}
\caption{ The scaled spectral density ${\tilde \chi}_{_V}(k)/(\phi_1\phi_2)$ versus wavenumber $k$ for disordered hyperuniform media for the
first three space dimensions, as obtained from relations (\ref{spec2-hyper}), (\ref{spec3-hyper}) and (\ref{spec1-hyper}). 
%Here we set the length scale $a$ in (\ref{auto-hyper}) to be unity.  
}
\label{spec-hyper}
\end{figure}

Figure \ref{spec-hyper} depicts the scaled spectral densities for the aforementioned disordered hyperuniform models
in the first three space dimensions. It is seen that the peak values increase substantially with increasing dimension.

\begin{figure}[bthp]
\centerline{\includegraphics[width=3.4in,keepaspectratio,clip=]{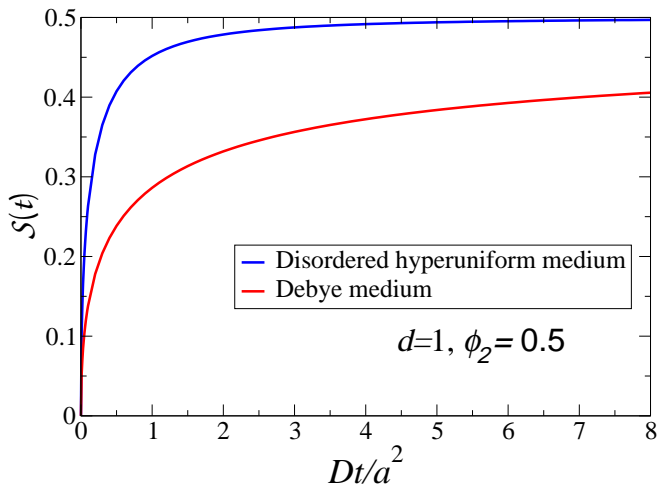}}
\caption{Comparison of the spreadabilities for Debye random media and disordered hyperuniform media for $d=1$ and $\phi_2=0.5$.  }
\label{compare}
\end{figure}

The $n$th moment $M_n(\chi_{_V})$ of the autocovariance function (\ref{auto-hyper}) for any $n$ 
is given exactly by 
\begin{equation}
M_n(\chi_{_V})= c \phi_1\phi_2  \frac{n!a^{n+1}}{[1+(qa)^2]^{n+1}} \left[ \cos(\beta)\cos(\theta) -\sin(\beta) \sin(\theta)\right],
\label{mom-hyper}
\end{equation}
where $\beta= (n+1)\arctan(qa)$.  The specific expressions for the moments for the parameters
used above for the first three space dimensions are given in Appendix \ref{mom}, which yield corresponding
exact representations of the  spreadability ${\cal S}(t)$ via the infinite series (\ref{6}).
Using these results and (\ref{hyp}) yields the corresponding long-time asymptotic expansions of ${\cal S}(t)$
for the first three space dimensions:
\begin{equation}
{\cal S}(\infty)- {\cal S}(t) = \frac{\phi_1}{4\sqrt{\pi} \,(Dt/a)^{3/2}} +{\cal O}\left((Dt/a)^{-5/2}\right) \quad (d=1),
\end{equation}
\begin{equation}
{\cal S}(\infty) -{\cal S}(t) = \frac{3\phi_1}{16\,  (Dt/a)^{2}} +{\cal O}\left((Dt/a)^{-3}\right) \quad (d=2),
\end{equation}
and
\begin{equation}
{\cal S}(\infty) -{\cal S}(t) = \frac{81\phi_1}{64\, \sqrt{\pi} (Dt/a)^{5/2}} +{\cal O}\left((Dt/a)^{-7/2}\right) \quad (d=3).
\end{equation}

For fixed dimension, we have already noted that the spreadability for disordered hyperuniform media 
will be substantially larger than that of nonhyperuniform media. Figure \ref{compare} specifically demonstrates
this distinction in one dimension by comparing the spreadabilities for Debye random media and disordered hyperuniform media.

\subsection{Antihyperuniform Media}

\begin{figure}[bthp]
\centerline{\includegraphics[width=3.4in,keepaspectratio,clip=]{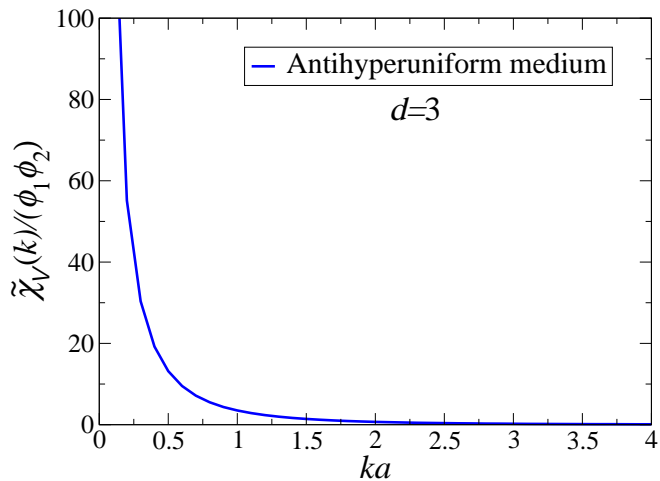}}
\caption{The scaled spectral density ${\tilde \chi}_{_V}(k)/(\phi_1\phi_2)$ versus wavenumber $k$ for antihyperuniform media 
in three dimensions, as obtained from (\ref{spec-anti}). 
%Here we set the length scale $a$ in (\ref{auto-anti}) to be unity.   
}
\label{anti-spec}
\end{figure}

As a model of antihyperuniform media in three dimensions, we consider here the following autocovariance function 
\begin{equation}
\frac{\chi_{_V}(r)}{\phi_1\phi_2} =\frac{1}{1+2(r/a) +(r/a)^2}.
\label{auto-anti}
\end{equation}
This monotonic functional form meets all of the known necessary realizability conditions on a valid autocovariance function \cite{To16b}.
It is clear that any $n$th order moment $M_n(\chi_{_V})$ for $n \ge 1$ is unbounded.  
The corresponding spectral density is given by
\begin{widetext}
\begin{equation}
{\tilde \chi}_{_V}(k)= \frac{4\pi a^2}{ka}  \Big[ \mbox{Ci}(ka)[ ka \cos(ka)+\sin(ka)]+\mbox{Ssi}(ka)[ka\sin(ka)-\cos(ka)\Big] ,
\label{spec-anti}
\end{equation}
\end{widetext}
where $\mbox{Ci}(x)\equiv \int_0^x dt \cos(t)/t$ is the cosine integral, $\mbox{Ssi}(x) \equiv \mbox{Si}(x) -\pi/2$ is the shifted sine integral
and $\mbox{Si}(x) \equiv \int_0^x dt \sin(t)/t$ is the sine integral.
We see that ${\tilde \chi}_{_V}(k)\sim2\pi^2/k$ in the limit $k\to 0$, which is consistent
with the power-law decay 
 $1/r^2$  of the $\chi_{_V}(r)$ in the limit $r \to \infty$.
The spectral density is plotted in Fig. \ref{anti-spec}.

We have already observed that the excess  spreadability for antihyperuniform media
will have the slowest decay to its infinite-time behavior relative to that of disordered hyperuniform media
or even to nonhyperuniform media in which the spectral density is bounded at the origin. 
These distinguished behaviors are clearly exhibited in Figure \ref{compare} where
the excess spreadabilities are compared for these three different cases
in three dimensions. The  long-time inverse power-law scalings of ${\cal S}(\infty)-{\cal S}(t)$
for the hyperuniform, nonhyperuniform and antihyperuniform three-dimensional  models 
are $1/t^{5/2}$, $1/t^{3/2}$ and $1/t$, respectively, as obtained from (\ref{long-time}).

\begin{figure}[bthp]
\centerline{\includegraphics[width=3.4in,keepaspectratio,clip=]{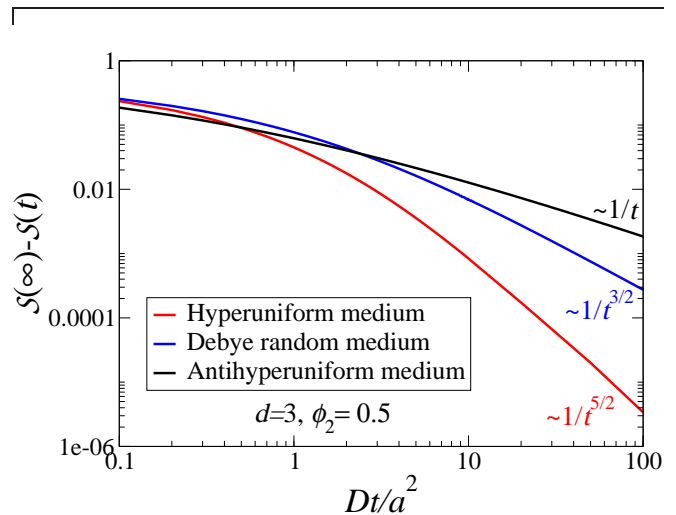}}
\caption{Comparison of the excess spreadabilities for Debye random media, disordered hyperuniform media and antihyperuniform  media for $d=3$ and $\phi_2=0.5$.  
The long-time inverse power-law scalings of ${\cal S}(\infty)-{\cal S}(t)$  for each of these models is  indicated. }
\end{figure}

\section{Applications to Stealthy Hyperuniform Media}
\label{app-stealthy}

In Sec. \ref{long}, we indicated that that the infinite-time aysmptotes of the spreadability of stealthy hyperuniform media
are approached exponentially fast and hence faster than any inverse power-law,
which applies to nonhyperuniform and nonstealthy hyperuniform media. In this section,
we explicitly demonstrate such long-time behaviors of both stealthy disordered and ordered media.
We also describe how the speadability of stealthy hyperuniform media
is linked to the covering problem of discrete geometry \cite{Co93,To10d}.

\subsection{Disordered Stealthy Hyperuniform Sphere Packings}

Consider a packing of identical spheres of radius $a$, which
we take to be phase 2. The packing fraction is $\phi_2=\rho v_1(a)$, where
$\rho$ is the number density and $v_1(a)$ is the volume of a sphere [cf. (\ref{v1})]. The spectral
density of such a packing, hyperuniform or not, can be expressed in terms of the structure factor $S({\bf k})$ according to \cite{To85b,To02a,To16b}
\begin{eqnarray}
{\tilde \chi}_{_V}({\bf k}) {\tilde \chi}({\bf k}) &=& \rho\, {\tilde m}^2(k;a) S({\bf k})  \nonumber \\
&=& \phi_2\, {\tilde \alpha}_2(k;a) S({\bf k})  
\label{chi-S}
\end{eqnarray}
where ${\tilde m}(k;a)$ is the Fourier transform of the sphere indicator function,
\begin{eqnarray}
{\tilde \alpha}(ka) &=& 
\frac{1}{v_1(a)} {\tilde m}^2(k;a) \nonumber \\
&=& \frac{1}{v_1(a)} \left(\frac{2\pi a}{k}\right)^{d} J_{d/2}^2(ka)  \nonumber \\
&=&  2^d \pi^{d/2} \Gamma(d/2+1)\frac{J_{d/2}^2(ka)}{k^d}.
\end{eqnarray}

\begin{equation}
{\tilde \alpha}_2(k;a) =  2^d \pi^{d/2} \Gamma(d/2+1)\frac{J_{d/2}^2(ka)}{k^d}
\end{equation}
is the Fourier transform of the scaled intersection volume of two spherical windows \cite{To03a}.
The zero-$k$ and large-$k$  of this function are given respectively by
\begin{equation}
{\tilde \alpha}_2(0;a) = v_1(a)
\end{equation}
and
\begin{equation}
{\tilde \alpha}_2(k;a) \sim 2^{d+1} \pi^{d/2-1} \Gamma(1+d/2)\frac{ \cos^2[ka -(d+1)/4]}{a k^{d+1}} \quad (ka \to \infty).
\label{asym}
\end{equation}
Moreover, we have the following integral condition:
\begin{equation}
\frac{1}{(2\pi)^d}\int_{\mathbb{R}^d} {\tilde \alpha}_2(k;a) d{\bf k}=1.
\end{equation}
If the point configuration specified by the sphere centers is hyperuniform, then $\lim_{|{\bf k}|\to 0}S({\bf k})=0$, and hence 
the dispersion or packing is hyperuniform, since it immediately follows from
(\ref{chi-S}) that the stealthy hyperuniformity condition (\ref{condition}) on the spectral density is obeyed. Moreover, if the sphere centers
constitute a stealthy and hyperuniform point configuration, $S({\bf k})=0$ for $ 0 \le |{\bf k}| \le K$, and hence
it follows that the spectral density is also identically zero up to the cut-off value $K$,
i.e., it obeys relation (\ref{stealth}).

Disordered stealthy hyperuniform packings have been generated using the collective-coordinate
optimization procedure \cite{To15} by decorating the resulting ground-state point configurations
by nonoverlapping spheres \cite{Zh16b,Ki20a}.  The degree of order of such ground states depends on a tuning parameter
$\chi$, which measures the extent to which the  ground  states  are  constrained by the size of the cut-off value
$K$ relative to the number of degrees of freedom. For $\chi <1/2$, the ground states are typically
disordered and uncountably infinitely degenerate in the infinite-volume limit \cite{To15}.
Using the fact that $\rho \chi=v_1(K)/[2d(2\pi)^d]$ \cite{To15}, it immediately follows
that for identical nonoverlapping spheres of radius $a$ that the dimensionless stealthy cut-off value
$Ka$ in terms of the packing fraction $\phi_2$ for any space dimension $d$ is given by
\begin{equation}
(Ka)^d=d 2^{d+1}\Gamma^2(1+d/2) \phi_2 \chi.
\label{K}
\end{equation}

Given the specific stealthy form obtained from (\ref{chi-S}), one
can compute the spreadability from formula (\ref{3}). Our main interest here is to determine from
this formula the exact long-time asymptotic form for disordered stealthy packings. Noting that
at long times, the spectral density can be replaced with its constant value at $k=K$, we find   
\begin{widetext}
\begin{equation}
{\cal S}(\infty)-{\cal S}(t) \sim \frac{d \omega_d}{(2\pi)^d } {\tilde \alpha}_2(Ka) S(K) \int_K^\infty k^{d-1} \exp(-k^2 D t) dk = \frac{d \omega_d}{2(2\pi)^d } {\tilde \alpha}_2(Ka) S(K) \frac{\exp(-K^2 Dt)}{K^2Dt} \; (Dt/a^2 \gg 1).
\label{S-stealth}
\end{equation}
\end{widetext}
We see that the decay of the excess spreadability  of a disordered stealthy hyperuniform two-phase medium
 is exponentially faster than that of any class I hyperuniform system
in which the exponent $\alpha>1$, specified by (\ref{power}), is bounded.

\subsection{Ordered Stealthy Hyperuniform Sphere Packings}
\label{periodic}

It is instructive to compare and contrast the spreadability
of disordered stealthy hyperuniform packings to that of their ordered stealthy hyperuniform counterparts. 
For this purpose, we consider identical nonverlapping spheres of radius $a$ centered on the sites
of a periodic lattice, which are stealthy and hyperuniform up to the first Bragg peak \cite{To15}. We begin by noting
that the structure factor of the sites of a  Bravais lattice in $\mathbb{R}^d$, excluding forward scattering, is given by
\begin{equation}
S({\bf k}) =\frac{(2\pi)^d}{v_c} \sum_{{\bf Q} \neq {\bf 0}} \delta({\bf k} -{\bf Q}),
\label{factor}
\end{equation}
where $v_c$ is the volume of a fundamental cell in direct space and $\bf Q$
denotes a reciprocal lattice (Bragg) vector. Substitution of (\ref{chi-S}) and (\ref{factor})
into (\ref{3}) yields
\begin{equation}
{\cal S}(\infty)-{\cal S}(t)=\phi_2 \sum_{{\bf Q} \neq {\bf 0}}   \frac{{\tilde \alpha}_2(|{\bf Q}|a)}{v_1(a)} \exp[-|{\bf Q}|^2 Dt].
\label{packing-b}
\end{equation}
Alternatively, we can recast this equation by employing the angular-averaged
structure factor $S(k)$, which is given by
\begin{equation}
S(k) =\frac{(2\pi)^d}{v_c} \sum_{n=1} \frac{Z(Q_n)}{s_1(Q_n)} \delta(k - Q_n),
\label{S-radial}
\end{equation}
where $Z(Q_n)$ is the coordination number at radial distance $Q_n$, $s_1(R)=d \pi^{d/2}R^{d-1}/\Gamma(1+d/2)$ is the surface area of $d$-dimensional
sphere of radius $R$,
and $\delta(k)$ is a radial Dirac-delta function. 

\begin{table*}[t]
  \caption{\label{1D} The scaled first Bragg peak $Q_1a$ for one-dimensional periodic packings of spheres (rods) of radius $a$ derived from common crystal structures in terms
  of the packing fraction $\phi_2$. The corresponding maximal packing fraction $\phi_2^{max}$ for each
  structure is also listed. In the case of a periodic packing with a an $n$-particle basis,
  $\eta$ is the dimensionless length of the fundamental cell in terms of the minimal nearest-neighbor distance
  and hence the maximal packing fraction $n/\eta$ is always less than or equal to unity. The packing with the largest value of $Q_1a$ is the one derived from the
  integer lattice $\mathbb{Z}$. }
%  \begin{center}
  \begin{tabular}{c|c|c}
  \hline
  Crystal Structure                 & $Q_1a$ & $\phi_2^{max}$          \\ \hline
  Integer lattice    ($\mathbb{Z}$)     & $\pi \phi_2$ & 1   \\
  Periodic with $n$-particle basis    & $\pi \phi_2/n$ & $n/\eta$       \\
  \hline
  \end{tabular}
%  \end{center}
  \end{table*}
  \begin{table*}[t]
  \caption{\label{2D}
  The scaled first Bragg peak $Q_1a$ (raised to the power 2) for two-dimensional periodic packings of spheres (circular disks)  of radius $a$ derived from common crystal structures in terms
  of the packing fraction $\phi_2$. The corresponding maximal packing fraction $\phi_2^{max}$ for each
  structure is also listed. The packing with the largest value of $Q_1a$ is the one derived from the
  triangular lattice $A_2\equiv A_2^*$.}
  \begin{center}
  \begin{tabular}{c|c|c}
    \hline
  Crystal Structure                               & $(Q_1a)^2$ & $\phi_2^{max}$\\ \hline
  Triangular lattice     ($A_2\equiv A_2^*$)      &$(8\pi/\sqrt{3})\phi_2=  (14.5103\ldots)\phi_2 $   & $\pi/\sqrt{12}=0.9068\ldots$ \\
  Square lattice ($\mathbb{Z}^2=\mathbb{Z}^2_*$)  & $(4\pi) \phi_2=(12.5663\ldots)\phi_2$ &  $\pi/4= 0.7853\ldots$ \\
  Honeycomb  crystal  ($\mbox{Dia}_2$)  & $(4\pi/\sqrt{3})\phi_2=  (7.2551\ldots)\phi_2 $  &      $\pi/(3\sqrt{3})=0.6045\ldots$   \\
  Kagom\'{e}  crystal ($\mbox{Kag}_2$)  & $[8\pi/(3\sqrt{3})]\phi_2=  4.8367\ldots)\phi_2$&       $3\pi/(8\sqrt{3})=0.6801\ldots$   \\
  \hline
  \end{tabular}
  \end{center}
  \end{table*}

  \begin{table*}
    \caption{\label{3D}
    The scaled first Bragg peak $Q_1a$ (raised to the power 3) for three-dimensional periodic packings of spheres  of radius $a$ derived from common crystal structures in terms
    of the packing fraction $\phi_2$. The corresponding maximal packing fraction $\phi_2^{max}$ for each
    structure is also listed. The packing with the largest value of $Q_1a$ is the one derived from the
    BCC lattice $D_3^*$.
    }
    \begin{center}
    \begin{tabular}{c|c|c}
      \hline
    Crystal Structure                               & $(Q_1a)^3$ &  $\phi_2^{max}$\\ \hline
    BCC lattice ($D_3^*$)      &$(6\sqrt{2}\pi^2)\phi_2=  (83.7463\ldots)\phi_2 $   & $\sqrt{3}\pi/8=0.6801\ldots$     \\
    FCC lattice ($D_3\equiv A_3$)            & $(9\sqrt{3}\pi^2/2)\phi_2=(76.9259\ldots)\phi_2$ &  $\pi/\sqrt{18}=0.7408\ldots$ \\
    HCP crystal               & $(8\sqrt{6}\pi^2/3)\phi_2=(64.4679\ldots)\phi_2$ &  $\pi/\sqrt{18}=0.7408\ldots$\\
    SC lattice  ($Z_3 \equiv Z_3^*$) & $ 6\pi^2\phi_2 =(59.2176\ldots)\phi_2$    &  $\pi/6=0.5235\ldots$ \\
    Simple hexagonal lattice  &  $ 3\sqrt{3} \pi^2\phi_2= (51.2839\ldots)\phi_2$ & $\pi/(3\sqrt{3})=0.6045\ldots$ \\
    Diamond  crystal ($\mbox{Dia}_3$)       & $(9\sqrt{3}\pi^2/4)\phi_2=(38.4629\ldots)\phi_2$ &   $\sqrt{3}\pi/16=0.3400\ldots$\\
    Pyrochlore crystal ($\mbox{Kag}_3$)         &  $(9\sqrt{3}\pi^2/8)\phi_2=(19.2314\ldots)\phi_2$ & $\sqrt{2} \pi/12=0.3702\ldots$    \\
    \hline
    \end{tabular}
    \end{center}
    \end{table*}
  
  \begin{table*}
  \caption{\label{4D}
  The scaled first Bragg peak $Q_1a$ (raised to the power 4) for four-dimensional periodic packings of spheres  of radius $a$ derived from common crystal structures in terms
  of the packing fraction $\phi_2$. The corresponding maximal packing fraction $\phi_2^{max}$ for each
  structure is also listed. The packing with the largest value of $Q_1a$ is the one derived from the
   four-dimensional checkerboard lattice $D_4 \equiv D_4^*$.}
  \begin{center}
  \begin{tabular}{c|c|c}
    \hline
  Crystal Structure                          & $(Q_1a)^4$ &  $\phi_2^{max}$       \\ \hline
  $D_4$ lattice         & $64\pi^2 \phi_2$& $\pi^2/16 =0.6168\ldots$   \\
  $\mathbb{Z}^4$ lattice         & $32\pi^2 \phi_2$& $\pi^2/16 =0.3084\ldots$   \\
  $\mbox{Dia}_4$ crystal         &$32\pi^2 \phi_2$ & $\sqrt{5}\pi^2/125=0.1765\ldots$   \\
  $\mbox{Kag}_4$ crystal          & $(64\pi^2/5) \phi_2$& $\sqrt{5}\pi^2/128=0.1724\ldots$   \\
    \hline
  \end{tabular}
  \end{center}
  \end{table*}

Now we recognize that expression (\ref{S-radial}) for $S(k)$ applies more generally
to {\it periodic packings} in which there are $N$ particles per fundamental cell, provided that $Z(Q_n)$ is interpreted to be the {\it expected} coordination number at radial distance $Q_n$.
Thus, for periodic packings, we have
\begin{equation}
{\cal S}(\infty)-{\cal S}(t)=\phi_2 \sum_{n=1}  Z(Q_n)  \frac{{\tilde \alpha}_2(Q_n\,a)}{v_1(a)} \exp[-Q_n^2 Dt],
\label{packing-radial}
\end{equation}
where the packing fraction is given by
\begin{equation}
\phi_2 =\frac{N v_1(a)}{v_c}.
\end{equation}
At large times, the first term in the sum of (\ref{packing-radial}) is the dominant contribution
and so
\begin{equation}
{\cal S}(\infty)-{\cal S}(t) \sim   \frac{\phi_2 Z(Q_1)\,{\tilde \alpha}_2(Q_1\,a)}{v_1(a)} \exp[-Q_1^2 Dt] \qquad (Dt/a^2 \gg 1),
\label{S-order}
\end{equation}
where $Q_1$ is the first (smallest positive) Bragg wavenumber. Result (\ref{S-order}), which is also a lower bound for all times, means that among all periodic packings of identical
spheres in $\mathbb{R}^d$ at a fixed packing fraction $\phi_2$, the
one with the largest first Bragg peak will have the fastest approach to the infinite-time behavior in space dimension $d$.
In dimensions one, two, three and four, these optimal packings for the spreadability correspond to the integer lattice $\mathbb{Z}$, triangular
lattice $A_2$, body-centered cubic (BCC) lattice $D_3^*$ (dual to the face-centered cubic (FCC) or checkerboard lattice $D_3$),
and the four-dimensional checkerboard lattice $D_4$ \cite{To15}, respectively. 
Tables I-IV list the scaled first Bragg peak $Q_1a$ raised to the power $d$ for some 
periodic sphere packings derived from commonly known periodic (crystal) point patterns in one, two, three, and four dimensions, respectively; see Appendix \ref{lattices}
for mathematical definitions. An exact expression for the spreadability for all times for 1D integer lattice  packings 
is given in Appendix \ref{1D-integer} and compared to spreadabilities of 1D models of disordered media.

We see that both long-time relations (\ref{S-stealth}) and (\ref{S-order}) for disordered and ordered stealthy packings, respectively,
involve exponential decay rates that are determined by the size of the stealthy cut-off value $Ka$,
which equals $Q_1a$ in the ordered case.   
Now, since stealthy disordered ground states must have values of $\chi$ less than 1/2,
any periodic packing with $\chi >1/2$ (see Ref. \cite{To15}) will have a larger cut-off value $Ka=Q_1a$,
according to  (\ref{K}) and hence faster spreadabilities. By the same token, the spreadability is slower for any periodic packing
with a value of $\chi$ smaller than that of a disordered stealthy packing. For example, the pyrochlore
crystal in three dimensions has a maximum $\chi$ value of $\chi=\pi/(4/\sqrt{12})=0.2267\ldots$ \cite{To15} and hence
any disordered stealthy packing with $\chi$ greater than the pyrochlore value
has a faster spreadability. This is to be contrasted with the optimal BCC structure with a maximal
value of $\chi=2\sqrt{2}\pi/9=0.9873\ldots$ \cite{To15}.

\subsection{Link to Covering Problem of Discrete Geometry}
\label{covering}

It should not go unnoticed that the point configurations corresponding to the optimal sphere packings for the spreadability
 are also the best {\it coverings} in the first four 
space dimensions \cite{To10d}.
The covering problem asks for the point configuration that minimizes the radius of overlapping spheres circumscribed around each of the points 
required to cover $d$-dimensional Euclidean space $\mathbb{R}^d$ \cite{Co93}. While the spreadability involves the ``covering" of space by non-uniform
concentration fields (as illustrated schematically in Fig. \ref{cartoon}), it is intuitively reasonable to conclude that decorations of the points of good coverings by
identical nonoverlapping spheres correspond to media with large spreadabilities. 
%Furthermore, in the first four dimensions, the best coverings are also the best {\it quantizers} and \cite{To10d}. 
Furthermore, it is interesting to note that the best coverings in the first four space dimensions
are also the best {\it quantizers} and  minimizers
of large-scale density fluctuations \cite{To10d}.

\subsection{Optimal Particle Shape for Spreadability}

Would a decoration of a stealthy and hyperuniform point configuration in $\mathbb{R}^d$
by nonoverlapping identical nonspherical particles yield spreadabilities
that are larger than that of their spherical counterparts? We conjecture that the decoration
of such an infinite point configuration by identical spheres possesses the largest
spreadability among all identical convex particles. While proving this conjecture is beyond
the scope of the present paper, the key arguments to support it rest on the fact
that the $d$-dimensional sphere is perfectly isotropic (i.e., possesses infinite-fold rotational symmetry)
and is the closed set with the minimal surface area to volume ratio, a consequence
of the isoperimetric inequality.

\section{Link of the Spreadability to  NMR  and Diffusion MRI Measurements }
\label{NMR}

NMR techniques provide noninvasive means to characterize the microstructure of fluid-saturated
porous media \cite{Br79,Mit92b,Mi93,Se94,Or02,We05,No14}. Here we identify a heretofore unknown relationship
between the spreadability ${\cal S}(t)$ and the NMR pulsed field
gradient spin-echo (PFGSE) amplitude ${\cal M}({\bf k},t)$   \cite{Mit92b}
as well as MRI-measured water diffusion in biological media \cite{No14}.

Consider a fluid-saturated porous medium,
which invariably contains paramagnetic impurities at the interface.
%One can probe the microstructure of such a porous medium using NMR techniques \cite{Mit92b,Mi93,Se94,Or02}.
In particular, one can extract microstructural information of the porous medium from the
PFGSE amplitude ${\cal M}({\bf q},t)$, which  depends on the wave vector $\bf q$ and time $t$ \cite{Mit92b,Mi93,Se94,Or02}. The PFGSE amplitude
 contains information on both the spectrum (eigenvalues) and eigenfunctions of the diffusion
operator, which are determined by the microstructure of the porous medium.
For statistically isotropic media, the time-dependent diffusion coefficient
$D(t)$ is directly obtained from the first
derivative of the  logarithm of ${\cal M}(q,t)$
with respect to the square of the wavenumber $q\equiv |\bf q|$, namely,
\begin{equation}
\lim_{q\to 0} -\frac{\partial \ln {\cal M}(q,t)}{\partial q^2}= {\cal D}(t)\,t,
\end{equation}
where ${\cal D}(t)$ is the effective time-dependent diffusion coefficient of the porous medium.
The long-time limit of ${\cal D}(t)$ is the static effective diffusion coefficient ${\cal D}_e$ \cite{To02a}. 

Mitra {\it et al.} \cite{Mit92b} proposed a simple phenomenological ansatz, based on an effective diffusion propagator, that relates 
the PFGSE amplitude ${\cal M}({\bf k},t)$ to the spectral density of the
porous medium. They showed that this approximation provides accurate estimates of ${\cal M}({\bf k},t)$ for both periodic and disordered microstructures.
Now we observe that setting the wave vector $\bf k$ to zero
in their formula (7) (up to a normalization parameter) gives, after simplification, the {\it total magnetization} as
a function of time, i.e.,
\begin{equation}
{\cal M}({\bf q=0},t)- \phi_2 = \frac{1}{(2\pi)^d \phi_2} \int {\tilde \chi}_{_V}({\bf k}) \exp[-k^2 {\cal D}(t) t]\; d{\bf k},
\label{nmr}
\end{equation}
where $\phi_2$ here is the porosity and ${\cal M}({\bf q}=0,t=0)=1$.
Comparing this infinite-wavelength formula to relation (\ref{3}) for the excess spreadability  ${\cal S}(\infty)-{\cal S}(t)$ reveals that 
they are very similar  to one another in functional form, except for the fact that the diffusion coefficient
appearing in (\ref{nmr}) is the effective time-dependent one. One can map the former
to the latter problem via the transformations $ {\cal S}(\infty)-{\cal S}(t) \rightarrow {\cal M}({\bf q=0},t)- \phi_2$ 
and $D \rightarrow {\cal D}(t)$.
Indeed, the total magnetization ${\cal M}({\bf q=0},t)$
shares many qualitative and quantitative features with the spreadability
function ${\cal S}(t)$. For example, it is known that for porous media with {\it perfectly absorbing}
interfaces, the short-time behavior   of ${\cal M}({\bf q=0},t)$ is of order $t^{1/2}$ and 
proportional to the specific surface $s$ \cite{Mi93}, which, as we noted in Sec. \ref{short}, is exactly
the case in the small-$t$ behavior of  the spreadability ${\cal S}(t)$. At long times, formula (\ref{nmr})
for the power-law scaling (\ref{power}) of the spectral density has the following asymptotic behavior:
\begin{widetext}
\begin{equation}
{\cal M}({\bf q}=0,t)- \phi_2 =  \frac{ B\, \Gamma((d+\alpha)/2)\,\phi_2}{2^d\,{\pi}^{d/2}\,\Gamma(d/2)\, ({\cal D}_et/a^2)^{(d+\alpha)/2}} 
+{o}\left(({\cal D}_e t/a^2)^{-(d+\alpha)/2}\right) \quad ({\cal D}_e t/a^2 \gg 1).
\label{long-time-2}
\end{equation}
\end{widetext}
This formula is identical to long-time formula (\ref{long-time}) for the excess spreadability when $D$ is replaced
by the static effective diffusion coefficient ${\cal D}_e$. This remarkable link between the two problems  
indicates that ${\cal S}(t)$ itself may serve as a simple figure of merit to gauge 
time-dependent diffusion processes in complex media and hence infer salient microstructural information
about heterogeneous media.

\begin{figure*}[bthp]
\includegraphics[width=5in,keepaspectratio,clip=]{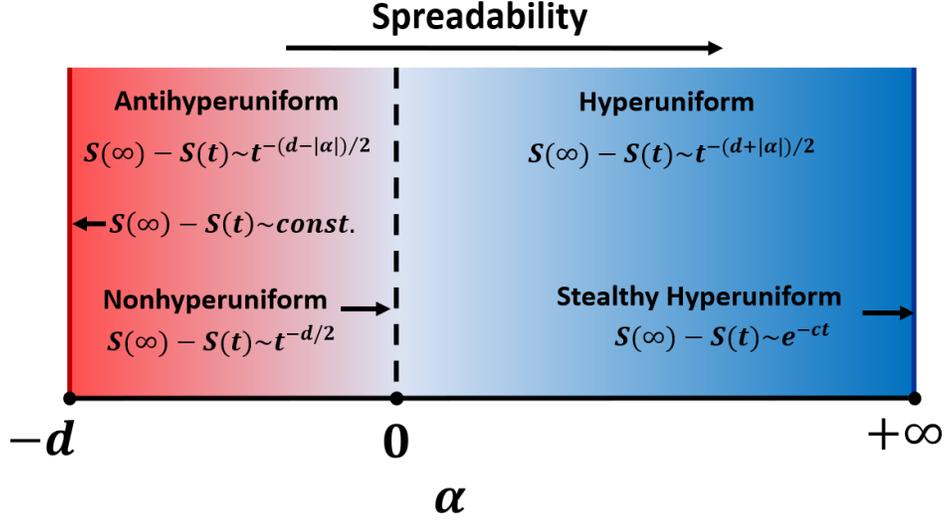}
\caption{``Phase diagram" that schematically shows the spectrum spreadability regimes in terms of the exponent $\alpha$.
As $\alpha$ increases from the extreme antihyperuniform limit of $\alpha \to -d$, the spreadability decay rate gets faster, i.e.,
the excess spreadability follows the inverse power law $1/t^{(d+\alpha)/2}$, except when $\alpha \to +\infty$, which
corresponds to stealthy hyperuniform media with a  decay rate that is exponentially fast.}  
\label{cartoon-2}
\end{figure*}

Diffusion-weighted magnetic resonance imaging (dMRI) has become a powerful tool for imaging
water-saturated biological media \cite{We05}. For the purpose of modeling water diffusion
in muscles and brain tissue, Novikov et al. \cite{No14} considered various one-dimensional models in which
diffusion is hindered by permeable barriers and estimated the corresponding long-time
behaviors of the time-dependent diffusion coefficient ${\cal D}(t)$. Based on this
one-dimensional analysis, they were able to extend their findings to any
space dimension and found the following long-time scaling behavior of $D(t)$:
\begin{equation}
{\cal D}(t)-{\cal D}_e \sim \frac{\mbox{C}}{t^{\varphi}},
\end{equation}
where $C$ is an undetermined structure-dependent constant and the exponent $\varphi=(d+\alpha)/2$.
Remarkably, we  see that the long-time behavior of ${\cal D}(t)-{\cal D}_e$ is identical to the excess
spreadability ${\cal S}(\infty)- {\cal S}(t)$, as specified
by the explicit scaling law (\ref{long-time}). While the spreadability problem is
substantially simpler than the determination of the effective time-dependent diffusion,
it is seen that, apart from constants, one can map the former
to the latter problem at long times via the transformations ${\cal S}(t) \rightarrow D(t)$ 
and ${\cal S}(\infty) \rightarrow {\cal D}_e={\cal D}(\infty)$.

\section{Discussion}
\label{discuss}

Our investigation has demonstrated  that the spreadability of diffusion information ${\cal S}(t)$ across time scales has the potential to serve as
a powerful dynamic figure of merit to probe and classify all translationally invariant two-phase microstructures 
across length scales. We established that the small-time behavior of ${\cal S}(t)$ is determined by 
the derivatives of the autocovariance function $\chi_{_V}({\bf r})$
at the origin, the leading  term of order $t^{1/2}$ being proportional to the specific surface $s$. We proved that the corresponding long-time behavior is determined by the form of the spectral density ${\tilde \chi_{_V}}({\bf k})$ at small wavenumbers, which enables one to ascertain the class of hyperuniform and nonhyperuniform media. 

In instances in which  the spectral density has the power-law form 
${\tilde \chi_{_V}}({\bf k})\sim |{\bf k}|^{\alpha}$ in the limit $|{\bf k}| \to 0$, the long-time excess spreadability 
for two-phase media in $\mathbb{R}^d$ is given by the following inverse power-law decay:
\begin{equation}
{\cal S}(\infty)-{\cal S}(t) \sim \frac{1}{t^{(d+\alpha)/2}}.
\label{S}
\end{equation}
Observe that this formula can distinguish among the possible strongest forms of hyperuniformity, i.e., class I, according
to the value of the exponent for any $\alpha >1$; the larger the value of $\alpha$ for such media, the faster the decay rate
the spreadability. The limit $\alpha \to +\infty$ corresponds to media in which the decay rate of ${\cal S}$ is faster than any inverse power law,
which we showed is the case for stealthy hyperuniform media. A measured long-time decay rate of ${\cal S}(\infty)-{\cal S}(t) \sim t^{-d/2}$, i.e., the case $\alpha=0$ in (\ref{S}),
would reveal a nonhyperuniform medium in which  the spectral density is a bounded positive number at the origin. On the other hand,
antihyperuniform media (with $-d < \alpha <0$) have the slowest decay among all
translationally invariant media, the slowest being when ${\cal S}(\infty)-{\cal S}(t)$ approaches a constant value (i.e., $\alpha \to -d$), independent of time. The stealthy hyperuniform
class is characterized by an excess spreadability with the fastest decay rate (exponentially fast)  among all hyperuniform media and hence
all translationally invariant microstructures. In short, the spreadability provides a dynamic means  
to classify the spectrum of possible microstructures that span between hyperuniform and nonhyperuniform media,
which is schematically illustrated in Figure \ref{cartoon-2}. Thus, in addition 
to the usual structure-based methods  to ascertain the hyperuniformity/nonhyperuniformity
of two-phase media discussed in Sec. \ref{class}, the spreadability at long times  provides
an alternative dynamic probe of such large-scale structural characteristics.

We obtained exact results for ${\cal S}(t)$ as a function of time for a variety of specific ordered and disordered model microstructures across dimensions, including antihyperuniform media, nonhyperuniform Debye random media,
nonstealthy hyperuniform media, disordered stealthy media and periodic media.
  We also demonstrated 
that the microstructures with ``fast" spreadabilities are also those that can be derived from efficient ``coverings" 
of Euclidean space $\mathbb{R}^d$.
%At first glance, it might appear that the conditions under which Prager's exact result are derived are not broadly applicable.
Finally, we  identified a remarkable connection between the spreadability  ${\cal S}(t)$ and noninvasive
nuclear magnetic resonance (NMR) relaxation measurements in physical and biological porous media \cite{Br79,Mit92b,Mi93,Se94,Or02,We05,No14}.

An interesting  avenue for future work is the generalization of the spreadability problem by relaxing Prager's assumption that the diffusion coefficients of both phases
are identical. This more general situation will involve expressions for  ${\cal S}(t)$ that now will not only 
involve the volume fractions and $S_2^{(i)}$, but all  higher-order correlation functions $S_3^{(i)},S_4^{(i)},\ldots$ as well
as the ratio of the phase diffusion coefficients. The solution of this general problem could be approached using a similar formalism as the ``strong-contrast"
methodology that has been developed to derive exact expressions for the  effective conductivity of two-phase media 
in terms of this infinite set of correlation functions and phase contrast ratio \cite{Se89,To02a}.

{\appendix

\section{Moments of the Autocovariance Function for the Disordered Hyperuniform Model}
\label{mom}

Here we provide simplified closed-form expressions obtained from the general formula (\ref{mom-hyper}) for the $n$th-order moment
of the autocovariance function (\ref{auto-hyper}) for the special cases in the first three
space dimensions considered in Sec. \ref{disorder-hyper}. Specifically, for $d=1$ with $qa=1$, and $c=\sqrt{2}$ and $\theta=\pi/4$,
we find 
\begin{equation}
M_n(\chi_{_V})=-\phi_1\phi_2 \frac{n!}{2^{n/2}} \sin(n\pi/4) \quad (d=1).
\end{equation}
Similarly, with $\theta=0$, $c=1$, we have for $d=2$ with $qa=1$,
\begin{equation}
M_n(\chi_{_V})=\phi_1\phi_2 \frac{n!}{2^{(n+1)/2}} \cos[(n+1) \pi/4] \quad (d=2)
\end{equation}
and for $d=3$ with $(qa)^2=1/3$,
\begin{equation}
M_n(\chi_{_V})=\phi_1\phi_2 \frac{n!3^{(n+1)/2}}{2^{n+1}} \cos[(n+1)\pi/6)] \quad (d=3).
\end{equation}

 \section{{Some $d$-Dimensional Crystal Structures}}
\label{lattices}
Here, we define some well-known crystal structures, including (Bravais) lattices
as well as lattices with a basis, what we generally call periodic point configurations \cite{To10d}.  
Some commonly known $d$-dimensional lattices include the hypercubic {$\mathbb{Z}^d$,
}{checkerboard} {$D_d$, and }{root} {$A_d$ lattices,
defined, respectively, by
}\begin{equation}{
\mathbb{Z}^d=\{(x_1,\ldots,x_d): x_i \in }{{\mathbb{ Z}}}{\} \quad \mbox{for}\; d\ge 1
}\end{equation}
\begin{equation}{
D_d=\{(x_1,\ldots,x_d)\in \mathbb{Z}^d: x_1+ \cdots +x_d ~~\mbox{even}\} \quad \mbox{for}\; d\ge 3
}\end{equation}
\begin{eqnarray}
A_d&=&\{(x_0,x_1,\ldots,x_d)\in \mathbb{Z}^{d+1}: x_0+ x_1+ \cdots +x_d =0\} \nonumber \\
&& \quad \mbox{for}\; d\ge 1,
\end{eqnarray}
{where $\mathbb{Z}$ is the set of integers ($\ldots -3,-2,-1,0,1,2,3\ldots$);
$x_1,\ldots,x_d$ denote the components of a lattice vector
of either $\mathbb{Z}^d$ or $D_d$; and $x_0,x_1,\ldots,x_d$ denote
a lattice vector of $A_d$. The $d$-dimensional lattices $\mathbb{Z}^d_*$, $D_d^*$ and $A_d^*$ are
the corresponding dual (or reciprocal) lattices.
Following Conway and Sloane \mbox{%DIFAUXCMD
\cite{Co93}
}%DIFAUXCMD
, we say that
two lattices are }{\it {equivalent}} {or }{\it {similar}} {if one becomes identical
to the other possibly by a rotation, reflection, and change of scale,
for which we use the symbol $\equiv$.
The $A_d$ and $D_d$ lattices can be regarded as $d$-dimensional generalizations
of the face-centered-cubic (FCC) lattice defined by $A_3 \equiv D_3$; however, for $d\ge 4$, they are no longer
equivalent. In two dimensions, $A_2 \equiv A_2^*$ defines the triangular lattice
with a dual lattice that is equivalent.
In three dimensions, $A_3^* \equiv D_3^*$ defines the body-centered-cubic (BCC)
lattice.
%DIF > A {\it self-dual} lattice $\Lambda$ is one with an {\it identical} dual lattice $\Lambda^*$ at density $\rho = \rho_* = 1/(2\pi)^{d/2}$, i.e., without any rotation,
%DIF > reflection, or change of scale for which we write $\Lambda=\Lambda^*$.\footnote{
%DIF > Mathematicians usually define a dual Bravais lattice to have a
%DIF > fundamental cell volume $v_{F^*} = 1/v_{F}$ [i.e., without the factor of
%DIF > $(2\pi)^d$], in which case self-duality is defined with respect to unit density; see Ref.~\cite{Co93}.}
In four dimensions, the checkerboard lattice and its dual are equivalent,
i.e., $D_4\equiv D_4^*$. The hypercubic lattice $\mathbb{Z}^d\equiv \mathbb{Z}^d_*$ and its dual
lattice are equivalent for all $d$. 

We denote by $\mbox{Dia}_d$ and $\mbox{Kag}_d$ the crystals that
are $d$-dimensional generalizations of the diamond and  kagom{\' e} crystals, respectively, for $d\ge 2$  \cite{Za11e}.
While the crystal $\mbox{Dia}_d$ has a two-particle basis (independent of $d$), the crystal $\mbox{Kag}_d$
as a ($d+1$)-particle basis.

\section{Spreadability for 1D Integer Lattice Packings}
\label{1D-integer}

%consider the hypercubic lattice $\mathbb{Z}_d$ with a lattice
%spacing of $L$ for which we know
%\begin{equation}
%S({\bf k}) =\frac{(2\pi)^d}{L^d} \sum_{{\bf q} \neq {\bf 0}}  \delta({\bf k} -{\bf q}),
%\label{cubic}
%\end{equation}
%where the $i$th component of $\bf q$, denoted by $q_i$, is given by
%\begin{equation}
%q_i = \frac{2 \pi n_i}{L}
%\end{equation}
Here we derive an exact expression for the spreadability for all times for the special case of one-dimensional  packings of identical rods
of radius $a$ (length $2a$) centered on the sites of the integer lattice $\mathbb{Z}$ with lattice
spacing $L$, so that $Q_n = 2 \pi n/L$ and $\phi_2=2a/L$. Application of the general formula (\ref{packing-radial})
in the case of the 1D integer lattice packing, where $Z(Q_n)=2$ for all $n$, yields
\begin{eqnarray}
{\cal S}(\infty)-{\cal S}(t)
&=&\phi_2 \sum_{n=1}^{\infty}   \frac{{\tilde \alpha}(2\pi n a/L)}{a} \exp[-(2\pi n)^2 Dt/L^2] \nonumber \\
&=&\frac{2}{\,\phi_2 \,\pi^2} \sum_{n=1}^{\infty}   \frac{\sin^2(\pi n \phi_2)}{ n^2 } \exp[-(2\pi n )^2 Dt/L^2]. \nonumber \\
\label{sum}
\end{eqnarray}
Note that because ${\cal S}(t=0)=0$, we have the identity
\begin{equation}
\frac{1}{\pi^2} \sum_{n=1}^{\infty}   \frac{\sin^2(\pi n \phi_2)}{n^2} =\frac{\phi_1 \phi_2}{2}.
\end{equation}
%and the inequality
%\begin{equation}
% \frac{\phi_1 \phi_2}{2} \ge \frac{1}{\pi^2} \sum_{n=1}^{\infty}   \frac{\sin^2[(\pi n \phi_2)]}{n^2}  \exp[-(2\pi n )^2 Dt/L^2] \quad \mbox{for all}\; t.
%\end{equation}
%Moreover,
%\begin{eqnarray}
%\beta(\infty)-\beta(t)
%&\ge &   \frac{2}{\,\phi_1 \,\pi^2}  \sum_{n=1}^{\infty} \exp[-(2\pi )^2 Dt/L^2]  \nonumber \\
%&=&\phi_2 \exp[-(2\pi )^2 Dt/L^2].
%\end{eqnarray}

\begin{figure}[bthp]
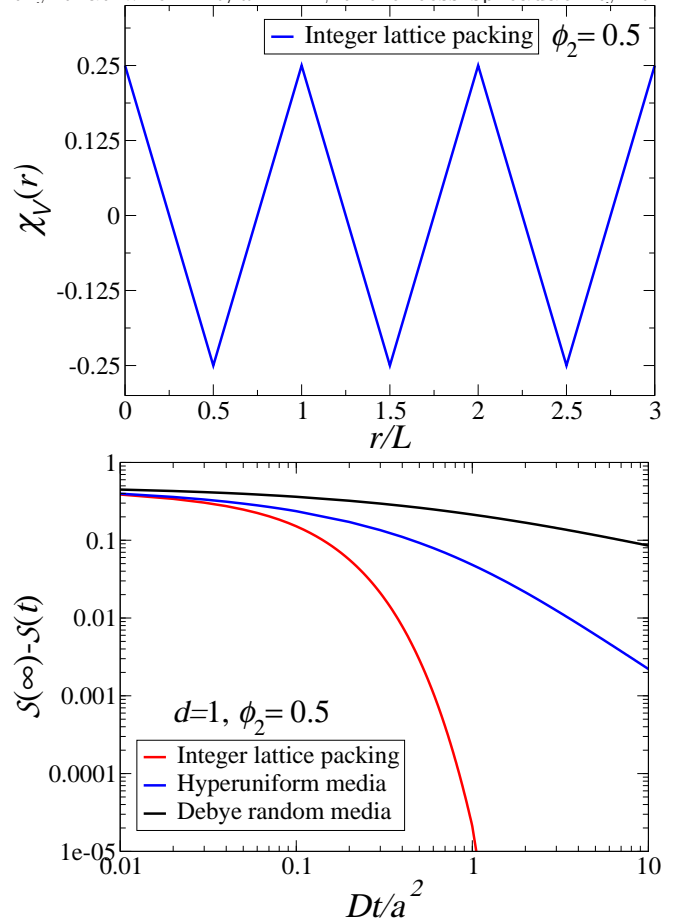

\includegraphics[width=3.4in,keepaspectratio,clip=]{autocov-integer-0.5.eps}\\
\includegraphics[width=3.4in,keepaspectratio,clip=]{integer.and.1D.eps}
\caption{ Top panel: The autocovariance function $\chi_{_V}(r)$ versus $r/L$
for the integer lattice packing for the instance $\phi_2=0.5$, where $L$ is the lattice spacing. Bottom panel: Excess spreadability 
${\cal S}(\infty)-{\cal S}(t)$ versus dimensionless time $D t/a^2$ for short times for three different models:
integer lattice packing, nonstealthy disordered hyperuniform
media and (nonhyperuniform) Debye random media, each with $\phi_2=0.5$. }
\label{int}
\end{figure}

The autocovariance function $\chi_{_V}(r)$ for the integer lattice packing for the instance $\phi_2$
and the corresponding excess spreadability for small times is shown in Fig. 1. The latter
plot compares ${\cal S}(\infty)-{\cal S}(t)$ to those of the 1D models of (nonhyperuniform) Debye random media
and nonstealthy disordered hyperuniform media, as  discussed in Sec. \ref{Debye} and Sec. \ref{disorder-hyper}, respectively.
It is noteworthy that when $Dt/a^2=1$, the excess spreadability for periodic media 
is already about four orders of magnitude smaller than that of nonstealthy disordered hyperuniform media.

}

\begin{acknowledgments}

The author thanks Jaeuk Kim, Michael Klatt, Murray Skolnick and Yang Jiao for very helpful discussions.
He is grateful to Michael Klatt for his assistance in creating Figures 1 and 2,
and Yang Jiao and Jaeuk Kim for their assistance in creating Figure 9.
Acknowledgment is made to the Donors of the American Chemical Society Petroleum Research Fund under
Grant No. 61199-ND9 for support  of this research.

\end{acknowledgments}

%\bibliography{new}

%merlin.mbs apsrev4-1.bst 2010-07-25 4.21a (PWD, AO, DPC) hacked
%Control: key (0)
%Control: author (0) dotless jnrlst
%Control: editor formatted (1) identically to author
%Control: production of article title (0) allowed
%Control: page (1) range
%Control: year (0) verbatim
%Control: production of eprint (0) enabled
%

\end{document}